\documentclass[useAMS,usenatbib]{mn2e}
\usepackage{natbib, graphicx, times}
\usepackage{amsmath}
\usepackage{mathtools}
\usepackage{array}
\usepackage{epstopdf}
\usepackage{amssymb}
\usepackage{extarrows}
\usepackage{color}
\usepackage{enumitem}
\usepackage{scrextend}
\usepackage{threeparttable}

\setlist{leftmargin=7.5mm}

\newcommand{\lsim}{\mathrel{\rlap{\lower4pt\hbox{\hskip1pt$\sim$}}
    \raise1pt\hbox{$<$}}}                
\newcommand{\gsim}{\mathrel{\rlap{\lower4pt\hbox{\hskip1pt$\sim$}}
    \raise1pt\hbox{$>$}}}                
   
\newcommand{\cbf}{} 

\def\arcsec{\hbox{$^{\hbox{\rlap{\hbox{\lower4pt\hbox{$\,\prime\prime$}}}\hbox{$\frown$}}}$}}

\title[Low-mass planet-induced vortices]{Which planets trigger longer-lived vortices: low-mass or high-mass?}

\author[Hammer, Lin, Kratter, \& Pinilla]{Michael Hammer$^{1}$\thanks{E-mail: mhammer@email.arizona.edu},
Min-Kai Lin$^{2, 3}$,
Kaitlin M. Kratter$^{1}$,
Paola Pinilla$^{4}$ \\
$^{1}$ Steward Observatory, University of Arizona, Tucson, AZ 85721, USA \\
$^{2}$ Institute of Astronomy and Astrophysics, Academia Sinica, Taipei 10617, Taiwan \\
$^{3}$ Physics Division, National Center for Theoretical Sciences, Taipei 10617, Taiwan \\
$^{4}$ Max Planck Institute for Astronomy, K\"{o}nigstuhl 17, 69117 Heidelberg, Germany
}

\begin{document}

\date{Accepted XXX. Received YYY; in original form ZZZ}

\pagerange{\pageref{firstpage}--\pageref{lastpage}} \pubyear{2016}

\maketitle

\label{firstpage}

\begin{abstract}
Recent ALMA observations have found many protoplanetary discs with rings that can be explained by gap-opening planets less massive than Jupiter. Meanwhile, recent studies have suggested that protoplanetary discs should have low levels of turbulence. Past computational work on low-viscosity discs has hinted that these two developments might not be self-consistent because even low-mass planets can be accompanied by vortices instead of conventional double rings. We investigate this potential discrepancy by conducting hydrodynamic simulations of growing planetary cores in discs with various aspect ratios ($H/r=0.04$, 0.06, 0.08) and viscosities ($1.5 \times 10^{-5} \lesssim \alpha \lesssim 3 \times 10^{-4}$), having these cores accrete their gas mass directly from the disc. With $\alpha < 10^{-4}$, we find that sub-Saturn-mass planets in discs with $H/r \le 0.06$ are more likely to be accompanied by dust asymmetries compared to Jupiter-mass planets because they can trigger several generations of vortices in succession. We also find that vortices with $H/r = 0.08$ survive $>6000$ planet orbits regardless of the planet mass or disc mass because they are less affected by the planet's spiral waves. We connect our results to observations and find that the outward migration of vortices with $H/r \ge 0.08$ may be able to explain the cavity in Oph IRS 48 or the two clumps in MWC 758. Lastly, we show that the lack of observed asymmetries in the disc population in Taurus is unexpected given the long asymmetry lifetimes in our low viscosity simulations ($\alpha \sim 2 \times 10^{-5}$), a discrepancy we suggest is due to these discs having higher viscosities.
\end{abstract}




\section{Introduction} \label{sec:intro}

It was once thought that protoplanetary discs needed to have high levels of turbulence to explain observed stellar accretion rates \citep[e.g.][]{alexander14, hartmann16} because it was assumed that turbulence drives angular momentum transport through the disc. This turbulence was thought to originate from hydrodynamic and magnetohydrodynamic (MHD) instabilities, most notably the magneto-rotational instability \citep[MRI:][]{balbus91, balbus98}. Recent computational studies, however, have demonstrated that non-ideal MHD effects should suppress the MRI in the bulk of the disc's midplane \citep{bai13a, bai17, wang19} because of the low levels of ionization in this region \citep{gammie96, fromang02, cleeves13}. 

Without the MRI, stellar accretion rates may instead be driven by magneto-thermal disc winds \citep{bai16}, while other instabilities could instead generate a weaker amount of turbulence \citep[e.g. the vertical shear instability:][]{nelson13, mkl15, manger20, flores20}. Observational studies have offered preliminary support for the idea that protoplanetary discs have low levels of turbulence. By analyzing spectral line broadening, \cite{flaherty15, flaherty18} were able to constrain the level of turbulence in certain discs to be $\alpha < 10^{-3}$ (where $\alpha$ is the standard turbulent parameter from \citealp{alpha}), which is weaker than expected in the surface layers of the disc. Meanwhile, \cite{rosotti20} demonstrated that the deviations from Keplerian velocity at the edges of planetary gaps in a few specific discs indicate the level of turbulence is weak if grain growth is also limited.

These low levels of turbulence provide favorable conditions for gap-opening planets to generate vortices through the Rossby Wave instability \citep[RWI:][]{lovelace99, li00, li01}, a global non-axisymmetric instability that can arise in regions of a disc with a maximum in their RWI critical function profile, which is proportional to the ratio of the density divided by the vorticity, a quantity also known as the inverse vortensity. Such maxima naturally occur at or near the edges of a gap opened by a planet where there is a maximum in the disc's density profile \citep{li05}. These maxima can become unstable and create a vortex if the bump is sufficiently sharp \citep{ono16} and the disc has sufficiently low viscosity of $\alpha < 10^{-3}$ \citep{deValBorro07}.

Despite this expectation that discs should have low viscosities, vortices are not prevalent in observations of protoplanetary discs. Vortex candidates in the form of large-scale crescent-shaped asymmetries only appear in $<25\%$ of resolved discs \citep{vanDerMarel21}. Furthermore, the fraction of conventional two-sided gaps with an adjacent crescent-shaped feature is even lower given that many discs contain multiple gaps, and that many observed asymmetries --- e.g. HD 142527 \citep{boehler17} --- are associated with cavities rather than gaps.



Why aren't more gaps associated with crescent-shaped asymmetries? One possibility is that not all of these gaps are associated with planets. A single planet can open multiple gaps \citep{dong17b}. Even without planets, it is possible to create gaps through other means such as MHD-driven zonal flows \citep{johansen09, riols20}. Although vortices can also form at the edges of these gaps \citep[e.g.][]{krapp18}, their properties have not yet been studied in detail. Regardless of the source of these gaps, disc self-gravity can inhibit the RWI altogether in more massive discs, even in discs that are not gravitationally unstable \citep{mkl11, mkl12b, lovelace13}. It has also been demonstrated that dust feedback can limit vortex lifetimes to a few hundred orbits in 2-D simulations \citep{fu14b}, albeit feedback in 3-D simulations has been shown to have less of an effect \citep{lyra18}. Layered turbulence in 3-D discs has also been shown to shorten vortex lifetimes \citep{mkl14}.

Even without those physical effects, \cite{fu14a} found that only planets well above Jupiter-mass in discs with a narrow range of aspect ratios near $H / r = 0.06$ trigger long-lived vortices that survive on the order of ten thousand planet orbits. \cite{hammer17} extended that work and showed that these lifetimes dropped significantly when incorporating the time it takes to grow a gas giant planet during the runaway gas growth phase of core accretion, a process that can last hundreds to thousands of orbits \citep{lissauer09}.

Could it just be that the planet candidates associated with gaps in ALMA observations are not massive enough to generate vortices? \cite{lodato19} collected a sample of 48 of these gap-opening planet candidates and estimated many of them to have mass ratios between Neptune-mass and Jupiter-mass relative to the Sun. The prevalence of planets in this mass range accentuates the need to study vortices induced by lower-mass planets. These planets have been neglected in previous studies even though it has been demonstrated that planets below Saturn-mass can trigger vortices in viscous discs \citep{hammer17, hallam20} and super-Earths can trigger vortices in inviscid discs \citep{fung17}.

In this work, we investigate the dependence of vortex lifetime on planet mass and disc aspect ratio, with a focus on planets less massive than Jupiter. We also develop a more realistic model for the planet's growth compared to our method from \cite{hammer17}. Additionally, we incorporate simulations of the dust in these discs to help measure vortex lifetimes and determine how vortices should appear in observations.

This work is organized as follows: We provide an overview of the hydrodynamic simulations and the associated methodology in Section~\ref{sec:methods}, and outline the parameters we explored in Section~\ref{sec:simulations}. We present an overview of our main results in Section~\ref{sec:results}, and a more in-depth analysis of specific cases in Section~\ref{sec:gas-results}. We then discuss our synthetic observations and their methodology in Section~\ref{sec:synthetic}, caveats in Section~\ref{sec:caveats}, and how our results apply to observations in Section~\ref{sec:applications}. We conclude our work in Section~\ref{sec:conclusions}.


\section{Methods} \label{sec:methods}

We use the hydrodynamic code FARGO3D (see Section~\ref{sec:code}) to conduct a parameter study of planetary cores accreting gas from the low viscosity discs in which they are embedded. We explore planets accreting gas at different rates, along with discs that have different aspect ratios and viscosities. The simulations in our main parameter study only include the gas component of the disc (see Sections~\ref{sec:one-hydro} and~\ref{sec:planet}). For each case, we also conduct a two-fluid simulation consisting of both a gas fluid and a dust fluid that represents a single fixed grain size to probe the end of the dust asymmetry lifetime and potentially other important phases in the evolution of the vortex. The majority of these two-fluid simulations start in the middle of the vortex lifetime (see Section~\ref{sec:two-hydro}).


\subsection{Hydrodynamic code} \label{sec:code}

FARGO3D \citep{FARGO3D, FARGO3D-dust} is a multi-purpose 3-D finite-difference Eulerian (magneto-)hydrodynamic grid code that is the successor to the original FARGO code \citep{FARGO}. 

Like its predecessor, FARGO3D can invoke the FARGO (Fast Advection in Rotating Gaseous Objects) algorithm to speed up simulations \citep{FARGO}. The algorithm relies on all of the grid cells in a given annulus having similar values for the azimuthal velocity, a routine property of any disc that rotates at close to a Keplerian rate. By subtracting out the average of this similar azimuthal velocity in each annulus when evaluating the Courant-Friedrich-Levy (CFL) condition, the code can increase the timestep by about an order of magnitude, making it feasible to run a suite of simulations with sufficient resolution for several thousand planet orbits.


\subsection{Gas dynamics} \label{sec:one-hydro}

\subsubsection{Hydrodynamic equations} \label{sec:equations}

In our study, we use the 2-D hydrodynamic setup of FARGO3D that solves the Navier-Stokes equations in cylindrical polar coordinates ($r$, $\phi$). These equations consist of the continuity equation, given by
\begin{equation} \label{eqn:continuity}
\frac{\partial \Sigma}{\partial t} + \vec \nabla \cdot (\Sigma \vec{v}) = 0,
\end{equation}
where $\Sigma$ is the gas surface density, $\vec v$ is the velocity vector, and $t$ is time; as well as the momentum equations given by
\begin{equation} \label{eqn:navier-stokes}
\Sigma \Big( \frac{\partial \vec{v}}{\partial t} + \vec{v} \cdot \vec  \nabla \vec{v} \Big) = - \vec \nabla P - \vec \nabla \phi + \nabla \cdot T
\end{equation}
where $P$ is the pressure and $\phi$ is the gravitational potential. The stress tensor $T$ is given by 
\begin{equation} \label{eqn:stress-tensor}
\vec T = \Sigma \nu \Big[ \vec{v} +  (\vec{v})^T - \frac{2}{3}(\vec \nabla \cdot \vec{v}) \vec{I} \Big],
\end{equation}
where $\nu$ is the kinematic viscosity of the disc and $\vec{I}$ is the identity tensor. We use a locally isothermal equation of state $P = \Sigma c_\mathrm{s}^2$, where $c_\mathrm{s} = H \Omega_\mathrm{K}$ is the sound speed, $H$ is the scale height, and $\Omega_\mathrm{K}$ is local Keplerian orbital frequency. We use a flat, constant aspect ratio $h \equiv H / r$ in all of our simulations. The scale height at the location of the planet is defined to be $H_0$. We also apply a constant viscosity of $\nu = \hat{\nu} r_\mathrm{p}^2 \Omega_\mathrm{p}$, where $r_\mathrm{p}$ is the planet's fixed orbital radius and $\Omega_\mathrm{p}$ is the planet's orbital frequency. For simplicity, we typically refer to $\hat{\nu}$ as $\nu$ in subsequent mentions. As all of the discs in our study are locally isothermal, our simulations do not need to solve the energy equation.

\subsubsection{Disc setup} \label{sec:gas-disc}

The disc in each simulation begins with a power-law radial surface density profile of $\Sigma = \Sigma_\mathrm{0} (r / r_\mathrm{p})^{-1}$, where $r_\mathrm{p}$ is the orbital radius of the planet. We choose the value of the initial surface density at the location of the planet $\Sigma_\mathrm{0} = 5.787 \times 10^{-4}~M_{\bigstar} r_\mathrm{p}^{-2}$ to set the total disc mass in the simulation domain to be $20~M_\mathrm{Jup}$. 
Although this disc mass is relatively high, we neglect self-gravity. We discuss the caveats of leaving out self-gravity as well as the dependence of our results on the disc mass in Section~\ref{sec:self-gravity}. The disc has a temperature profile that follows $T(r) \propto r^{-1}$, consistent with its constant aspect ratio. The simulation domain extends from $r \in [0.2, 5.7]r_\mathrm{p}$ in radius and $\phi \in [0, 2 \pi]$ in azimuth. The grid across this domain consists of $N_\mathrm{r} \times N_\mathrm{\phi} = 1536 \times 2048$ arithmetically-spaced cells. This radial spacing resolves the scale height at the location of the planet and in the outer disc by at least 11 cells for $h = 0.04$, at least 16 cells for $h = 0.06$, and at least 22 cells for $h = 0.08$.

We employ evanescent boundary conditions towards the inner edge ($1.00~~r_\mathrm{in}<~r~<~1.25~r_\mathrm{in}$) and outer edge ($0.84~r_\mathrm{out}~<r~<~1.00~r_\mathrm{out}$) of every simulation. These boundary conditions are designed to damp waves by reducing the density and velocity fields towards their initial values \citep[e.g.][]{deValBorro06}. We keep the outer zone mostly unperturbed by rapidly damping any changes on a timescale of $1/500$th the orbital period at the outermost radius in the simulation domain. The inner zone is only damped on a timescale of 1/3rd the orbital period at the innermost radius. We employ standard periodic boundary conditions in the azimuthal direction.

\subsection{Planet} \label{sec:planet}

\subsubsection{Planet setup} \label{sec:planet-setup}

We begin each simulation with a $0.05~M_\mathrm{Jup}$ planet core at $r = r_\mathrm{p}$ on a fixed circular orbit around a solar-mass star ($M_{\bigstar} = 1.0~M_{\odot}$). The planet orbits at an angular frequency of $\Omega_\mathrm{p} = \sqrt{GM_\mathrm{\odot} / r_\mathrm{p}^3}$, where $G$ is the gravitational constant. For convenience, we set $r_\mathrm{p} = \Omega_\mathrm{p} = G = M_\mathrm{*} = 1$ in each simulation. As such, the planet completes one orbital period in a time of $T_p = 2\pi$. We model the planet's gravitational potential as 
\begin{equation} \label{eqn:potential}
\phi_\mathrm{p}(\mathbf{r}) = \frac{GM_\mathrm{p}}{\sqrt{(\mathbf{r} - {\mathbf{r}_\mathrm{\mathbf{p}}})^2 + r_\mathrm{s}^2}}
\end{equation}
where a smoothing length of $r_\mathrm{s} = 0.6H_0$ is included to avoid a gravitational singularity. The planet's initial core mass is introduced over $T_\mathrm{growth} = 5~T_\mathrm{p}$ following a profile of $m_\mathrm{p}(t) = \sin^2{\left(\pi t / 2T_\mathrm{growth}\right)}$. The planet is allowed to accrete from the start of the simulation.


\subsubsection{Accretion onto the planet} \label{sec:accretion}

Our main interest in modelling accretion onto the planet is to have the gap edges evolve self-consistently with the growth of the planet so that the properties of the resulting vortices are more realistic. We emphasize that we are not attempting to model the accretion process onto the planet in great detail. The actual accretion process is difficult to resolve in 2-D global hydrodynamic simulations and requires local or mesh-refined 3-D simulations to study \citep[e.g.][]{ayliffe09, szulagyi17a}. 

We begin each simulation with a $0.05~M_\mathrm{Jup}$ planetary core that can accrete gas from the disc within a fraction of its Hill sphere following the scheme used in the original FARGO code, where the Hill radius is defined as $R_\mathrm{H} \equiv a(q/3)^{1/3}$, $a$ is the planet semimajor axis, and $q$ is the planet-to-star mass ratio. This scheme is based on \cite{kley99} and contains two similar steps that are performed at each timestep. First, a fraction $f_1$ of the mass contained within all grid cells less than $0.75~R_{H}$ from the planet is removed. Then, a larger fraction $f_2$ of the mass in all grid cells less than $0.45~R_{H}$ from the planet is also removed. The values of the fractions are $f_1~=~1/3 \times A \times \Omega_\mathrm{p} dt$ and $f_2~=~2/3 \times A \times \Omega_\mathrm{p} dt$, where the timestep $dt$ is included to scale the fractional removals of mass $1/3 \times A$ and $2/3 \times A$ to take place over an amount of time $\Delta t$ = 1 instead of instantaneously. Whereas the accretion parameter $A = 1$ by default in the original FARGO code, we instead vary $A$ to model planets growing at different rates \citep[analogous to the methods used by][]{muley19, bergez20}.

This accretion scheme is intended to model the growth of the planet from a solid core to a gas giant during the runaway gas accretion phase of the core accretion model when the planet rapidly accretes the vast majority of its atmosphere. 
This phase begins at about when the atmosphere's mass surpasses the core mass \cite[e.g.][]{dangelo10}. We neglect the time it takes for the atmosphere's mass to reach the core mass primarily because this earlier phase can last for hundreds of thousands of years or longer, which would greatly increase the time needed to run each simulation and also be unlikely to affect the outcome.

We therefore chose a relatively high initial planet mass of $0.05~M_\mathrm{Jup}$, equivalent to $\approx 16~M_{\oplus}$ (where $M_\mathrm{Jup}$ is the mass of Jupiter and $M_{\oplus}$ is the mass of the Earth) so that our core could be thought of as already having accreted some of its atmosphere. As a result, we are starting out with a core mass that is larger than the presumed masses for gas giant cores, which are typically around $\approx 10~M_{\oplus}$. This choice also speeds up the time it takes for the lowest-mass planets to trigger the RWI.

The rate at which the planet accretes is almost always less than the physical limit set by the mass flux through the cross-section of the planet where it can accrete gas. The planet's cross section itself is set by the minimum of the Hill radius ($R_\mathrm{H}$) and the Bondi radius ($R_\mathrm{B}$), the latter of which is defined as $R_\mathrm{B} \equiv 2GM_\mathrm{p}/c_\mathrm{s}^2$, where $G$ is the gravitational constant, $M_\mathrm{p}$ is the planet mass, and $c_\mathrm{s}$ is the sound speed \citep[e.g.][]{dangelo10}. The physical limit on the accretion rate is then
\begin{equation}
\dot{M}_\mathrm{p,limit} = \frac{\Sigma \Omega_\mathrm{p}}{\sqrt{2 \pi} H} \left(\min \{R_\mathrm{H}, R_\mathrm{B} \}\right)^2.
\end{equation}
For giant planets, the minimum radius is the Hill radius, as the Bondi radius becomes the larger of the two at the relatively low crossover mass of
\begin{equation}
M_\mathrm{p,cross} = 0.044~M_\mathrm{Jup} \left(\frac{h}{0.06} \right)^{3},
\end{equation}
where $R$ is the separation from the star. With the high starting core mass and the planet accreting from only a fraction of the Hill sphere in our simulations, the planet naturally accretes at a rate below the physical limit except at the beginning of simulations with $h = 0.08$ where the planet has $M_\mathrm{p,cross} = 0.104~M_\mathrm{Jup}$. The planet's initially unphysical accretion rate is not a concern in this case since the planet does not trigger the RWI until after it grows past the crossover mass.


\subsection{Dust dynamics} \label{sec:two-hydro}

\subsubsection{Overview} \label{sec:overview}

We also use FARGO3D to simulate dust with the gas, where the dust is implemented as a pressureless fluid \citep{FARGO3D-dust}. The dust also experiences radial and azimuthal drag forces as a result of being coupled to the gas that cause it to behave differently from the gas. These drag forces act as additional source terms added to the momentum equation given by
\begin{equation} \label{eqn:rad_drift}
\frac{\partial v_\mathrm{r, d}}{\partial t} = - \frac{v_\mathrm{r, d} - v_\mathrm{r}}{t_\mathrm{s}},
\end{equation}
\begin{equation} \label{eqn:az_drift}
\frac{\partial v_\mathrm{\phi, d}}{\partial t} = - \frac{v_\mathrm{\phi, d} - v_\mathrm{\theta}}{t_\mathrm{s}},
\end{equation}
where $v_\mathrm{r}$ and $v_\mathrm{\theta}$ are the radial and azimuthal velocity components and $v_\mathrm{d}$ and $v$ are the dust and gas components respectively. We only simulate small particles where the stopping time  falls in the Epstein regime \citep{weidenschilling77}. The stopping time in the midplane is defined as 
\begin{equation} \label{eqn:stopping}
t_\mathrm{s} = \frac{\mathrm{St}}{\Omega} = \left( \frac{\pi}{2} \frac{\rho_\mathrm{d} s}{\Sigma} \right) \frac{1}{\Omega},
\end{equation}
where $\rho_\mathrm{d}$ is the physical density of each dust grain (which we take to be 1 g / cm$^3$) and $s$ is the size of each grain. The Stokes number St is the dimensionless form of the stopping time. 

In addition to drag forces, the dust experiences turbulence from the gas that can cause it to diffuse through the disc. The dust diffusion is modeled as another source term \citep{clarke88} included in the continuity equation as
\begin{equation} \label{eqn:diffusion}
\frac{\partial \Sigma_\mathrm{d}}{\partial t} = \nabla \cdot \left( D \Sigma \nabla\left( \frac{\Sigma_\mathrm{d}}{\Sigma}\right) \right),
\end{equation}
where the diffusion coefficient $D = \hat{D} r_\mathrm{p}^2 \Omega_\mathrm{p}$ and $\hat{D} \approx \hat{\nu}$ since the particles in this study are small enough to satisfy St$~\ll~1$ \citep{youdin07}. We note that due to the low viscosities in our study, incorporating dust diffusion has hardly any effect on the behavior of the dust in our simulations. Additionally, we defer incorporating dust feedback to our future work and discuss the implications of leaving it out in this study in Section~\ref{sec:dust-feedback}.



\subsubsection{Dust setup} \label{sec:dust-simulations}

For each gas simulation in our main parameter study, we run a corresponding simulation of a single-size dust grain primarily to determine the end of the vortex lifetime. To conserve computational resources, we initialize these two-fluid gas and dust simulations in the middle of the gas simulation towards the end of the vortex lifetime. 

Similar to the gas, the dust is initialized to a matching power-law radial surface density profile of $\Sigma_\mathrm{d} = \Sigma_\mathrm{d, 0} (r / r_\mathrm{p})^{-1}$ with a constant initial dust-to-gas ratio of $\Sigma_\mathrm{d, 0} / \Sigma_\mathrm{0}$ = 1/100. The dust spans the same simulation domain as the gas and uses the same boundary conditions. We simulate dust grains with a reference initial Stokes number of St$_0$ = 0.023 at the location of the planet. This Stokes number corresponds to a grain size of
\begin{equation} \label{eqn:st}
s = 1.9~\mathrm{mm} \left(\frac{\rho_\mathrm{d}}{1~\mathrm{g/cm^3}} \right)^{-1} \left(\frac{M_{_*}}{1~M_{\odot}} \right) \left(\frac{r_\mathrm{p}}{20~\mathrm{AU}} \right)^{-2},
\end{equation}
depending on the physical separation between the star and the planet, the mass of the star, and the physical density of each dust grain. The grain size itself is fixed throughout the simulation domain, leaving the Stokes number to vary as the gas surface density changes. We choose a Stokes number of St$_0$ = 0.023 to probe the vortex lifetimes because we found in our previous study that it lies in-between smaller grains that spread throughout most of the vortex and larger grains that are routinely much more concentrated \citep{hammer19}.

We initialized the two-fluid simulations at around when the gas vortex spreads out into a ring with the goal of selecting a start time of a few hundred to one thousand planet orbits before the dust asymmetry also spreads into a ring. Since the smooth axisymmetric initial conditions for the dust are not consistent with the gas component that already contains a vortex by the time we restart the simulation, we only used start times where the dust evolves into a clear asymmetry at the outer gap edge, a process that takes about 50 to 100 orbits. If there was no dust asymmetry by that point (in other words, if the vortex was already dead), we tried an earlier start time instead. We tested the effect of choosing different start times in some cases and found that it did not change the dust asymmetry lifetime by more than 100 orbits.


\subsubsection{Determining dust asymmetry lifetimes} \label{sec:signature}

We define the total dust asymmetry lifetime as the time from when the $m = 1$ vortex first appears in the gas until when it disappears in the dust either by fading into a ring by being replaced a different signature, and provided that the vortex stays at $r < 2.0~r_\mathrm{p}$. With this definition, we are actually measuring the lifetime of the dust asymmetry associated with the vortex, not the lifetime of the gas vortex itself. While it is typically straightforward to tell when the dust asymmetry forms just by looking for the appearance of the RWI in the gas, it is non-trivial to identify the end of the dust asymmetry lifetime just by analyzing the gas.

To find the end of the dust asymmetry lifetime, we monitor the standard deviation in the azimuthal dust profile in the vicinity of the vortex. We first radially-average the dust surface density across two scale heights around the peak in the radial surface density profile. Once the normalized standard deviation $\sigma(\Sigma_\mathrm{d}(\phi)) / <\Sigma_\mathrm{d}(\phi)>$ drops below a threshold of $10\%$, we consider the vortex dead. Because this quantity drops rapidly as the dust becomes more symmetric and spreads out into a ring, choosing a different threshold would not change the lifetime by more than 100 orbits. In cases where the RWI starts out as an $m = 1$ mode, we also use this method to identify that start of the lifetime.

Even though many cases have gas vortices that re-form, we note that the dust asymmetry persists in-between successive gas vortices. We did not find any cases where the dust asymmetry fades into a ring and then reappears later. In cases with $h = 0.08$, we mark the end of the vortex lifetime when the vortex migrates out to $r > 2.0~r_\mathrm{p}$. This definition excludes the phase when there are multiple vortices in the outer disc as well as the next phase when there is just a single vortex at a much larger separation than usual. We made this choice because the original vortex has moved much further away from the planet and because each of these vortices that migrated outwards survives past the end of the simulation.

We stress that the variation in the gas is not so useful for identifying the end of the dust asymmetry lifetime because the dust asymmetry does not disappear at the same as the gas asymmetry. Even when a vortex spreads into a ring, the outer gap edge retains a weak azimuthal pressure bump that can still trap dust even if the variation in the gas is very low.


\section{Suite of Simulations} \label{sec:simulations}


\subsection{Parameter Study} \label{sec:parameter-study}

\begin{table}
\caption{Parameter study. The final mass $M_\mathrm{p,~3000}$ is recorded at 3000 planet orbits. Planets in simulations that were extended beyond this point have higher mass by the end.}
\begin{tabular}{ c c l | l }
  $h$ & $\nu~/~r_\mathrm{p}^2 \Omega_\mathrm{p}$ & $A$ & $M_\mathrm{p,~3000} / M_\mathrm{J} $ \\
   \hline
        \hline
  0.08 & $10^{-7}$ & 0.01, 0.02, 0.05, 0.167 & 0.18, 0.33, 0.59, 1.22  \\
  0.06 & $10^{-7}$ & 0.02, 0.05, 0.167, 0.5 & 0.21, 0.36, 0.63, 0.92 \\
  0.04 & $10^{-7}$ & 0.05, 0.167, 0.5, 1.0 & 0.21, 0.36, 0.53, 0.67 \\
          \\
          \hline
          \\
  0.06 & $10^{-6}$ & 0.05, 0.167, 0.5 & 0.37, 0.70, 1.26 \\
  0.08 & $10^{-6}$ & 0.02, 0.05, 0.167 & 0.52, 0.84, 1.18  \\
          
\end{tabular}
\label{table:simulations}
\end{table}

We conduct a total of 18 one-fluid gas simulations in our main parameter study comprising three disc aspect ratios ($h=$ 0.04, 0.06, and 0.08), two different viscosities ($\nu = 10^{-6}$ and $10^{-7}$), and four different accretion parameters. These parameters are outlined in Table~\ref{table:simulations}. 

Twelve of the simulations use the lower viscosity $\nu = 10^{-7}$ (equivalent to $1.5 \times 10^{-5} \le \alpha \le 6 \times 10^{-5}$ at $r = r_\mathrm{p}$). For the intermediate scale height $h = 0.06$, we choose $A =$ 0.02, 0.05, 0.167, and 0.5. With these accretion parameters, the final planet masses at $t = 3000$ planet orbits fall in the range of $0.2~M_\mathrm{J}$ to $1.0~M_\mathrm{J}$ (where $M_\mathrm{J} \equiv M_\mathrm{Jup}$) with the lower viscosity $\nu = 10^{-7}$. For the lowest aspect ratio $h=0.04$, we replace $A = 0.02$ with $A = 1$. For the largest aspect ratio $h=0.08$, we replace $A = 0.5$ with $A = 0.01$. These replacements were chosen so that the range of final planet masses is similar for all three aspect ratios. We note that it is necessary to decrease the largest $A$ with the largest aspect ratio $h = 0.08$ because the gap-opening timescales are slower at a given planet mass. The opposite is true with the smallest aspect ratio $h = 0.04$. Lastly, the remaining six simulations use the larger viscosity $\nu = 10^{-6}$ (equivalent to $1.5 \times 10^{-4} \le \alpha \le 3 \times 10^{-4}$ at $r = r_\mathrm{p}$), only encompassing the two larger aspect ratios and their three largest respective accretion parameters.


\subsection{Additional Simulations} \label{sec:additional-simulations}

Outside of our main parameter study, we run additional simulations primarily to test the causes of various phenomena described in our results. These additional tests include simulations with different disc masses, planet masses, or viscosities not covered in our main parameter study. 
All of these supplementary simulations are described alongside the results {\cbf and caveats}.
We also present resolution tests in Appendix~\ref{sec:resolution}.


\section{Main results} \label{sec:results}


\subsection{Overview} \label{sec:overview}

\begin{figure} 
\centering
\includegraphics[width=0.47\textwidth]{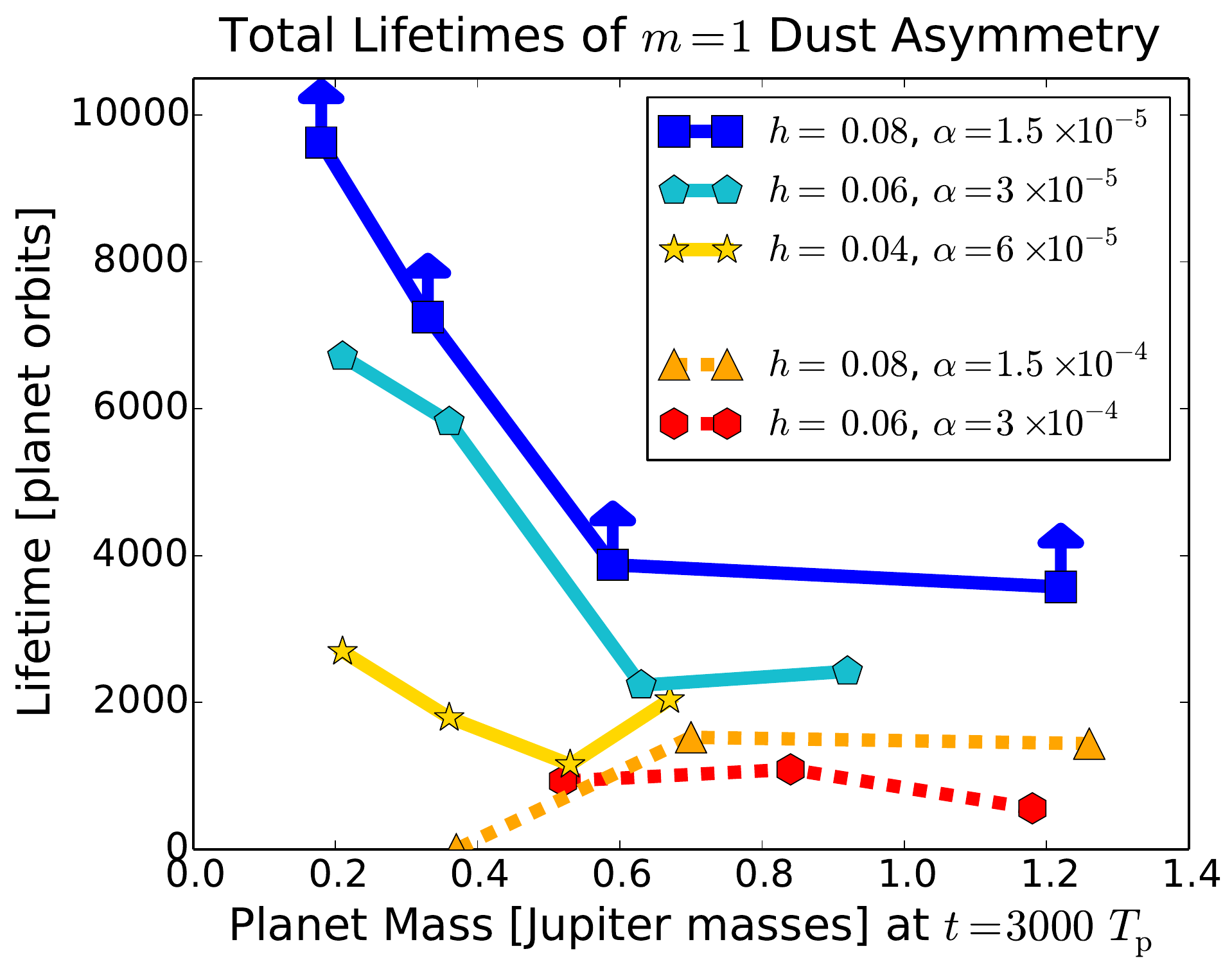}
\caption{Total $m = 1$ dust asymmetry lifetimes as a function of planet mass (recorded at $t = 3000$ planet orbits). With a low viscosity of $\nu = 10^{-7}$, vortices live longer when triggered by lower-mass planets than higher-mass ones. With $h = 0.08$ and  $\nu = 10^{-7}$, an asymmetry survives past the end of each simulation in all cases (as indicated by the arrows).} 
\label{fig:lifetimes}
\end{figure}

In discs with a low viscosity ($\nu = 10^{-7}$), we find that lower-mass planets and planets in discs with larger aspect ratios in general preferentially induce longer-lived dust asymmetries than higher-mass planets (see Figure~\ref{fig:lifetimes}). Similar trends occur in the cases with the higher viscosity ($\nu = 10^{-6}$), but are much weaker.

There are several different reasons that these dust asymmetries have longer lifetimes in the low viscosity cases with lower planet masses. With lower disc aspect ratios ($h \le 0.06$), the total lifetimes in these cases are longer even though the initial gas vortices are very short-lived. These initial vortices are almost all short-lived regardless of planet mass because of the combination of two effects: [1] the disc's viscosity halting the growth of the RWI after $\sim 10^3$ orbits, and [2] shocks from the planet's spiral waves distorting the vortex's smooth elliptical structure. Nonetheless, after the initial vortex spreads into a ring, those spiral shocks create a new maximum in the RWI critical function (defined in Section~\ref{sec:RWI}), making it possible to form later-generation vortices in every case. In the cases with higher-mass planets, however, the width of the gap grows too quickly for the location of the pressure bump to stay near the more inward location of the bump in the RWI critical function, preventing later-generation vortices from forming after only about $\sim 10^3$ orbits. On the contrary with lower-mass planets, the pressure bump stays closer to the planet and the RWI critical function maximum for a much longer period of time. As a result, a series of later-generation vortices can form that drastically extend the total lifetime of the dust asymmetry. We find that the cutoff between the low-mass and high-mass planet regimes is roughly the classical thermal mass $M_\mathrm{th} = 3 h^3 M_\mathrm{\bigstar}$, the mass at which a planet can open a ``full gap" with $R_\mathrm{H} \ge H_0$ \citep{lin93}.\footnote{Other studies define $M_\mathrm{th}$ with a coefficient of 1 or $\frac{2}{3}$ instead of 3 \citep{goodman01}.}

With larger disc aspect ratios ($h \ge 0.08$), the initial gas vortices are all very long-lived regardless of planet mass or disc mass. In our parameter study, the continued growth of the planet enables the RWI to grow unimpeded until the vortex becomes strong enough to shed angular momentum via its own spiral waves. With a lower disc mass, the deeper gap opened because of the accretion onto the planet halts the growth of the RWI like in the cases with lower disc aspect ratios. Nonetheless, the vortices in these cases still survive well over 6000 planet orbits because the planet generates much weaker spiral shocks in these cases.

\subsection{Theoretical Interpretation} \label{sec:theory}


\subsubsection{General background on RWI}  \label{sec:RWI}

Planets can generate vortices through the Rossby wave instability at their inner and outer gap edges where they create sharp maxima in the disc's radial inverse vortensity profile, where the vortensity is defined as the disc's density divided by its vorticity. The complete critical function for the RWI in an adiabatic disc is presented in \cite{lovelace99} as 
\begin{equation} \label{eqn:maximum-adiabatic}
L = \frac{\Sigma}{\omega}\mathcal{S}^{2/ \gamma},
\end{equation}
where $\omega \equiv (\nabla \times \mathbf{v})_\mathrm{z}$ is the vorticity, $\mathcal{S} \equiv P / \Sigma^{\gamma}$ is the entropy, and $\gamma$ is the adiabatic index. A similar critical function applies for locally isothermal discs and is given in \cite{mkl12a} as
\begin{equation} \label{eqn:maximum-iso}
L_\mathrm{iso} =  c_\mathrm{s}^2 \frac{\Sigma}{\omega}.
\end{equation} 
Because this function depends on both density and vorticity, the outer radial position at which the disc initially becomes unstable is located slightly interior to the peak in the disc's radial density profile by up to one scale height. 

The instability itself can be thought of as the interaction between two separate Rossby waves that form on either side of the maximum in the critical function \citep{umurhan10}. One Rossby wave forms on each side of the bump where there is a sharp vorticity gradient, moving in opposite directions along the azimuth. If the bump is narrow enough, these waves will interact and merge after growing by a sufficient amount in the radial direction. This merger makes the waves unstable and results in the formation of one or more vortices. Once these initial one or more vortices form, we observe that the growth of the instability in both density and vorticity causes the vortices to migrate outward slightly into the center of the pressure bump. At this point, the maximum in the critical function largely settles at the center of the pressure bump, allowing the vortices to continue to grow, albeit at a much slower rate.

We find that the location at which the initial maximum in the critical function $L_\mathrm{iso}$ arises depends strongly on the disc aspect ratio and only weakly on the planet mass. As the planet grows, its spiral waves generate a change in vortensity $[\omega/\Sigma]$ as they shock the disc given by 
\begin{equation}
\left[\frac{\omega}{\Sigma}\right] = - \frac{(M^2 - 1)^2}{\Sigma M^4} \frac{\partial u_\mathrm{\bot}}{\partial S} - \left(\frac{M^2 - 1}{\Sigma M^2 u_\mathrm{\bot}} \right) \frac{\partial c_\mathrm{s}^2}{\partial S},
\end{equation}
where $M = u_\mathrm{\bot} / c_\mathrm{s}$ is the Mach number, $u_\mathrm{\bot}$ is the pre-shock relative velocity perpendicular to the shock front, $S$ is the direction along the shock front away from the planet, and the first term involving $\partial u_\mathrm{\bot}/\partial S$ dominates the vortensity change \citep{li05, mkl10}. Previous work has shown that spiral shocks impart a positive change in vortensity in the horseshoe region close to the planet, leading up to a peak positive vortensity change occurring just outside of that region \citep[see Figure 6 in][]{mkl10}. Just beyond that peak, the planet imparts a negative change in vortensity in the rest of the outer disc. The peak negative vortensity change -- in other words, where the maximum in $L_\mathrm{iso}$ develops -- occurs only slightly exterior to the location at which the local Rossby number turns negative (where the local Rossby number is defined as $\mathrm{Ro} \equiv [\nabla \times (\mathbf{v} - {\mathbf{v}_\mathrm{\mathbf{K}}})]_\mathrm{z} / 2 \Omega$), and then decays at larger radii.

\subsubsection{Initial vortex formation}

\begin{figure*} 
\centering
\includegraphics[width=0.335\textwidth]{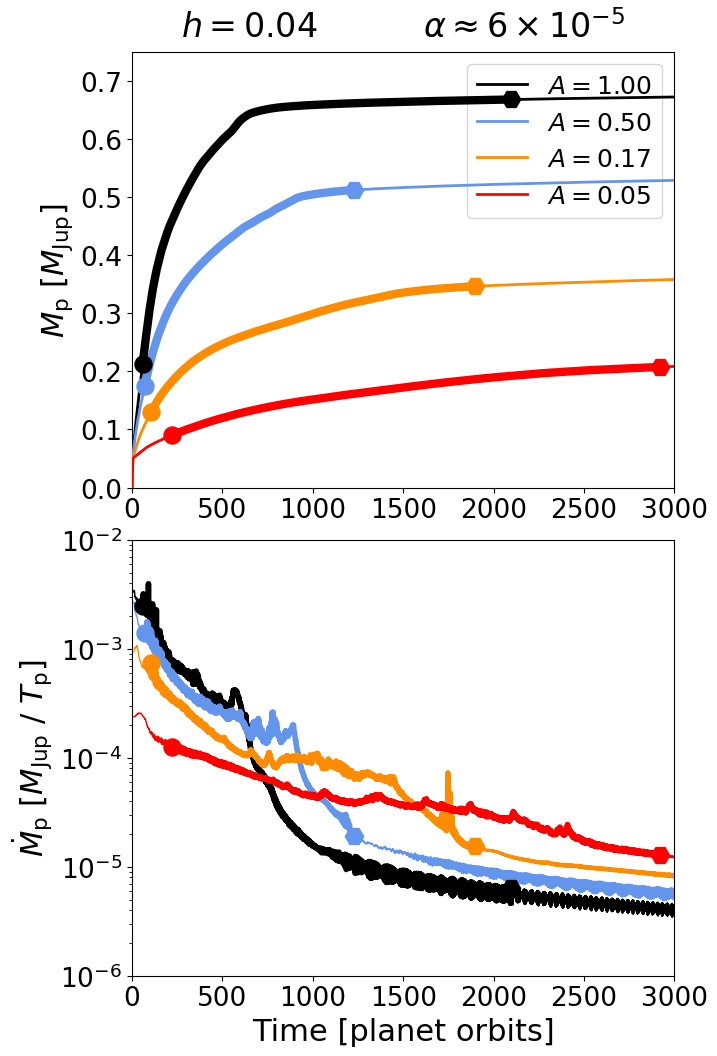}
\includegraphics[width=0.32\textwidth]{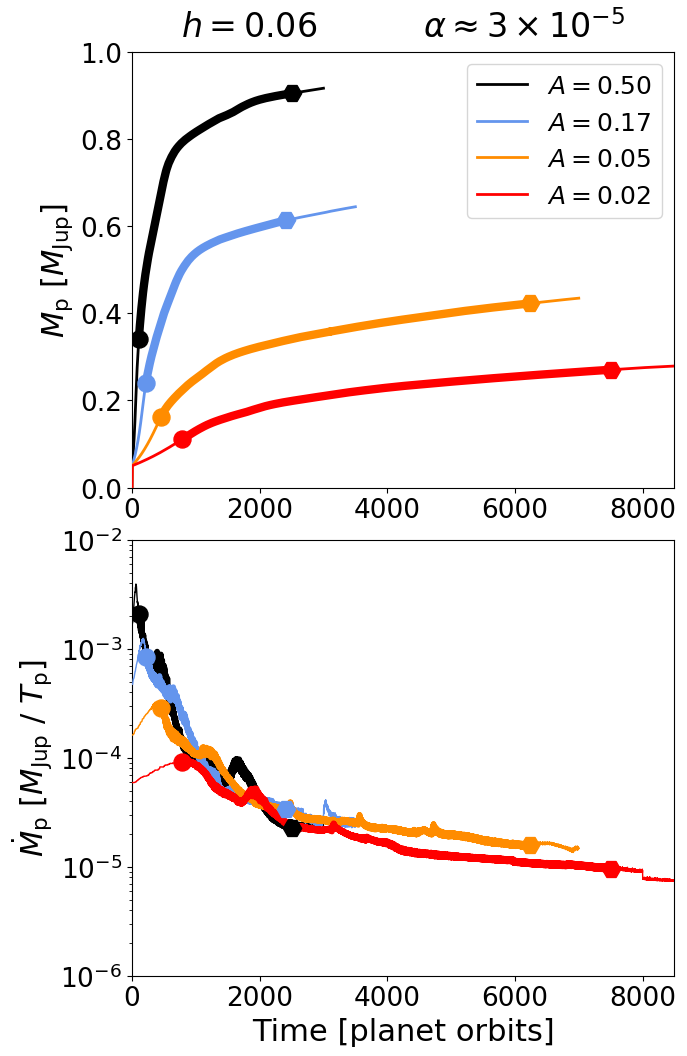}
\includegraphics[width=0.32\textwidth]{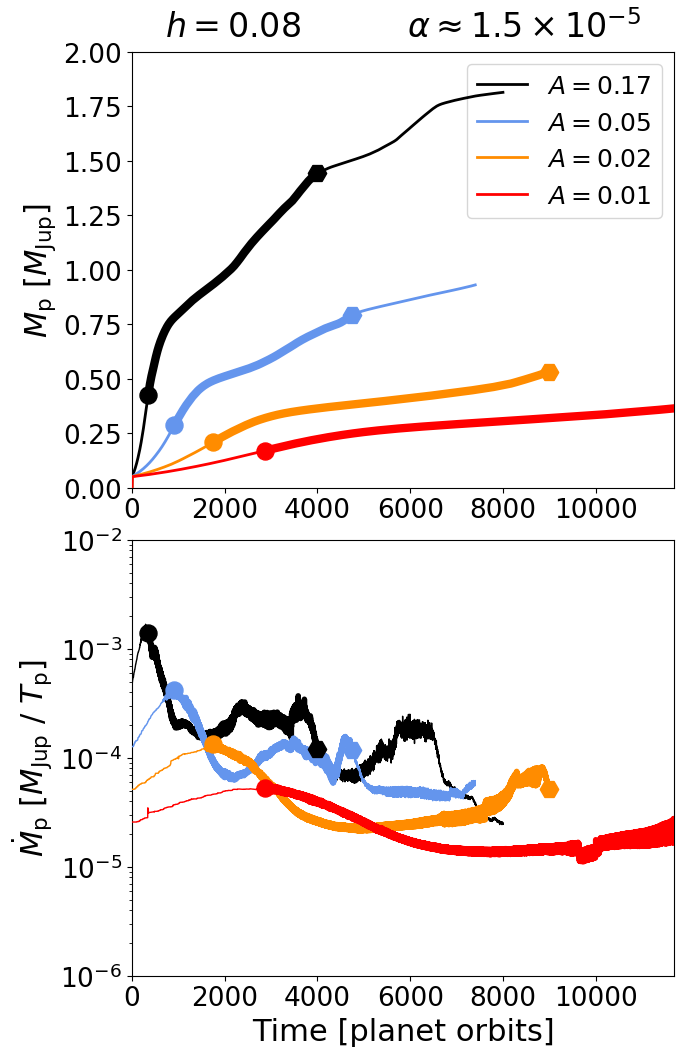}
\caption{Planet mass (\textit{top panels}) and planet mass accretion rate (\textit{bottom panels}) as a function of time for cases with the lower viscosity $\nu = 10^{-7}$.  A circle denotes the start of the vortex lifetime; a hexagon denotes the end. The vortex lifetime is highlighted in bold. The start and end of the vortex lifetime are marked on each side of the bold line with a circle and a hexagon respectively. We measure the accretion rates as the increase in planet mass over each orbit divided by one orbit.} 
\label{fig:mass}
\end{figure*}

\begin{figure*} 
\centering
\includegraphics[width=0.33\textwidth]{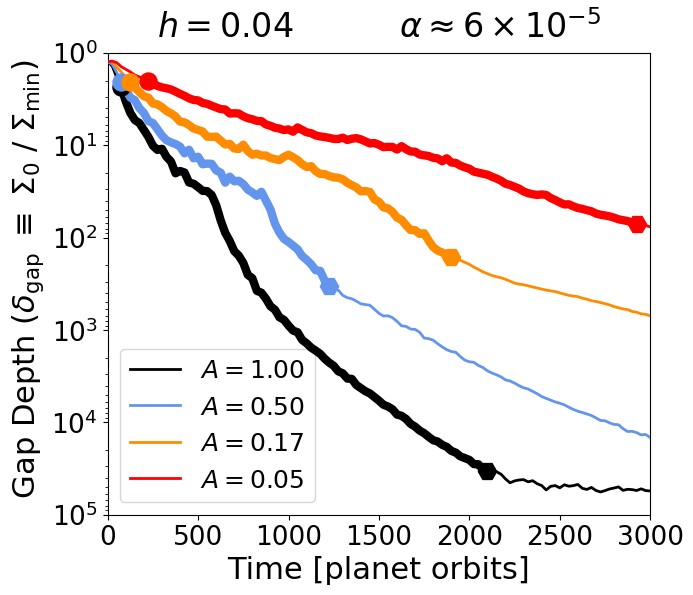}
\includegraphics[width=0.32\textwidth]{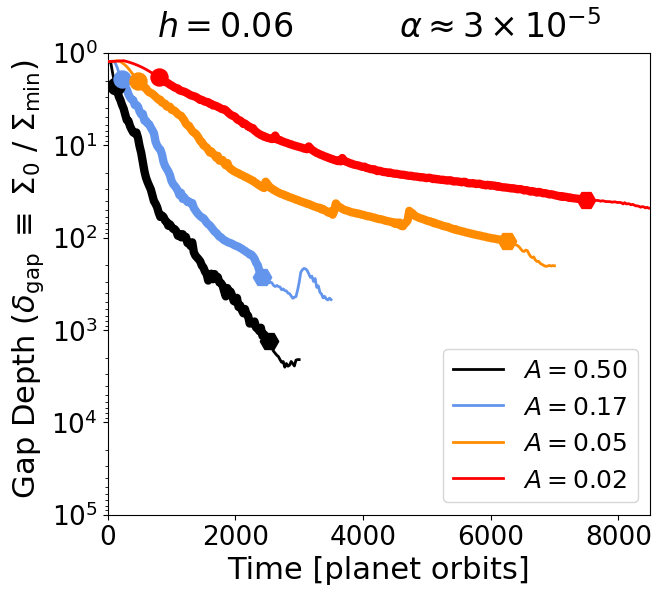}
\includegraphics[width=0.32\textwidth]{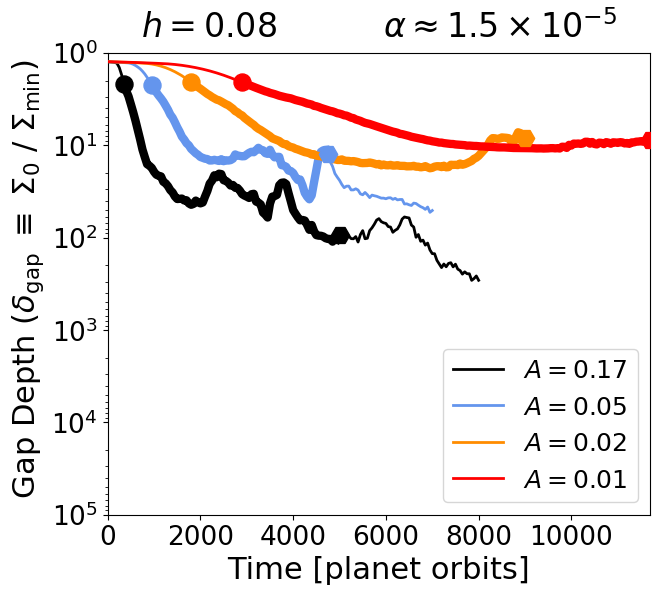}
\caption{Gap depths as a function of time for cases with the lower viscosity $\nu = 10^{-7}$. The vortex lifetime is highlighted in bold. The start and end of the vortex lifetime are marked on each side of the bold line with a circle and a hexagon respectively.} 
\label{fig:gap_depths}
\end{figure*}

Past computational studies of vortices that largely neglect the growth of the planet have found that the RWI generates compact vortices at the outer gap edge \citep[e.g.][]{fu14a}. These vortices have strong enough vorticity minima ($\mathrm{Ro} < -0.15$) to be well-described by the compact Gaussian vortex model developed by \cite{surville15}. In these simulations with planets growing to their final mass in $<100$ orbits, the RWI starts out with three to five small-scale co-orbital vortices as a result of one of the $m = 3$ to 5 modes beginning as the fastest-growing mode \citep{meheut12b}. Like the final $m = 1$ vortex that eventually emerges, these initial vortices also start out with strong vorticity minima ($\mathrm{Ro} < -0.15$). As a result, when the initial mode grows enough for the vortices to interact, they seamlessly merge in both their density and vorticity structures, creating a final $m = 1$ vortex with a compact overall structure.

On the other hand, the slower-growing planets in our study all generate vortices with elongated shapes, albeit through two separate pathways. Like the compact vortex cases from past studies, the faster-growing planets ($M_\mathrm{p,~3000} > M_\mathrm{th}$) in our parameter study also start out by generating three small-scale vortices. Unlike those compact cases, however, these initial vortices do not develop strong vorticity minima. As a result, when these vortices grow enough to interact, they only merge smoothly in their density structure while still remaining separated in their vorticity structure (see Section~\ref{sec:h06_super}). During this phase, the vorticity minima at the center of each initial vortex stay separate and rotate around each other. This behavior ultimately prevents the final $m = 1$ vortex from developing a vorticity minima at its center and leads to the vortex spreading out in azimuth instead of staying compact. The elongated vortex that develops is better described by the so-called GNG model from \cite{GNG} rather than the more compact model from \cite{surville15}.

With the slowest-growing planets ($M_\mathrm{p,3000} < M_\mathrm{th}$), the outer gap edge becomes unstable later on in the simulation ($>500~T_\mathrm{p}$) and these cases instead begin with an $m = 2$ or an $m = 1$ mode. These mode numbers arise because they have the fastest growth rates in their respective cases. The dominant mode numbers $m$ being lower in these cases indicates that the corresponding outer gap edges are closer to marginal instability \citep{ono16}, which is consistent with these cases having the slowest planet growth times. In the cases that start with an $m = 1$ mode, the vortex begins elongated with a weak vorticity minimum ($\mathrm{Ro} > -0.15$) and typically remains that way for the rest of its lifetime. In the larger disc aspect ratio ($h = 0.08$) cases in our parameter study, however, the vortex eventually becomes compact, but only after a few to tens of thousands of orbits. 

Regardless of the pathway, the elongated shape of a vortex does not necessarily lead to shorter total dust asymmetry lifetimes like in \cite{hammer17}. Planets in the sub-thermal mass regime generate vortices that sustain long-lived dust asymmetries ($> 3000~T_\mathrm{p}$) even though they start out elongated.


\subsubsection{Role of gap-opening process and timescales} \label{sec:gap}

The slow gap-opening timescales for low-mass planets in low-viscosity discs determine if and when each planet initially triggers the RWI, and can also sustain the growth of the RWI if the gap remains shallow.

We note that the starting planetary core mass in these cases is essentially already massive enough to eventually open up gaps that can trigger vortices. {\cbf Figure~\ref{fig:mass} shows the growth tracks --- that is, the planet masses and accretion rates over time --- for the planets in our study with $\nu = 10^{-7}$}. According to \cite{duffell13}, the gap-opening mass for planets in low-viscosity discs is given empirically by
\begin{equation} \label{eqn:true-gap-opening-mass}
M_\mathrm{gap} = h^{3/2} \sqrt{29~\delta_\mathrm{gap} \nu},
\end{equation}
where $\delta_\mathrm{gap} \equiv \Sigma_{0} / \Sigma_\mathrm{min}$ is the depth of the gap to be opened. For $\delta_\mathrm{gap} = 2$, the approximate gap depth at which all of the outer gap edges have already become unstable to the RWI in our simulations, $M_\mathrm{gap}~=~0.019$, 0.035, and 0.054 $M_\mathrm{Jup}$ for $h = $ 0.04, 0.06, and 0.08 respectively, implying that the starting core mass of $0.05~M_\mathrm{Jup}$ or just above should eventually trigger a vortex in the low viscosity cases.

Although the starting cores are already massive enough to open gaps that make the disc unstable, they do not open these gaps immediately. Figure~\ref{fig:gap_depths} shows the evolution of the gap depths for all of the $\nu = 10^{-7}$ cases. The gap depth never reaches a steady state over the timescales we explore, and in most cases it continually increases over time. Several different studies have shown that the gap depth depends on the planet mass, the disc aspect ratio, and the disc viscosity in the following way:
\begin{equation}
\delta_\mathrm{gap} - 1 \equiv \frac{\Sigma_{0}}{\Sigma_\mathrm{min}} - 1 = f_0 \frac{q^2}{h^3 \nu},
\end{equation}
where $q = M_\mathrm{p} / M_{\bigstar}$ and $f_0 = 0.45 / 3 \pi$ \citep{duffell13, fung14, kanagawa15}. For instance, a Saturn-mass planet ($0.3~M_\mathrm{Jup}$) should open a gap with $\delta_\mathrm{gap} = 84$ in a disc with $h = 0.08$. Yet, the case with $A = 0.02$ illustrates that the gap depth is only $\delta_\mathrm{gap} = 3.70$ when the planet reaches $0.3~M_\mathrm{Jup}$ at $t \approx 2600~T_\mathrm{p}$. This behavior is not surprising given that the time for a gap to reach a steady state can be approximated by the viscous timescale $t_\mathrm{visc} \sim h^2 / \nu$, which is inherently large due to the low viscosities associated with planet-induced vortices. For example with $\nu = 10^{-7}$, this timescale exceeds $10^4$ orbits and is longer than any of our simulations.

With these slow gap-opening timescales, the planet continues to grow substantially for roughly 500 to 2000 planet orbits in all cases. This growth helps sustain the growth of the RWI over the same timeframe as well until it is stopped by the disc viscosity. In the cases with a larger disc aspect ratio ($h = 0.08$) where the gap-opening timescales are the slowest, the gap remains shallow enough for the continued growth of the planet to prevent the viscosity from ever stopping the growth of the instability. This outcome requires a relatively high disc mass and/or a disc aspect ratio even larger than $h = 0.08$. Otherwise with a lower disc mass, the accretion onto the planet itself depletes the gap enough to prevent the planet and in turn the vortex from continuing to grow significantly.


\subsubsection{Role of later-generation vortices} \label{sec:later}

\begin{figure} 
\centering
\includegraphics[width=0.43\textwidth]{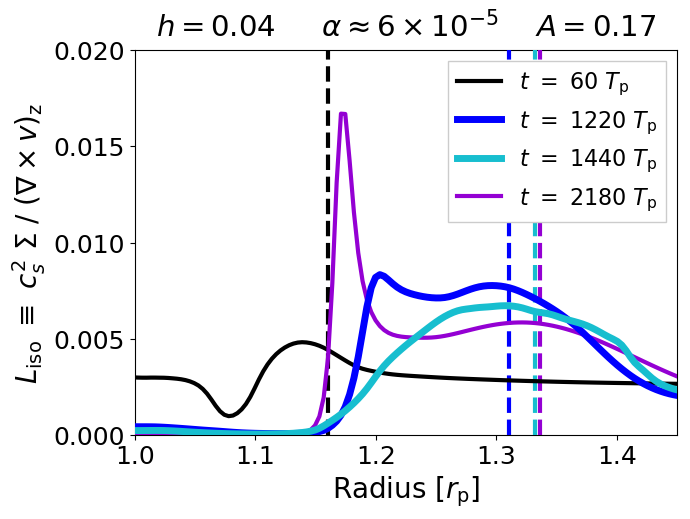} \\
\caption{Example radial profiles of the RWI critical function $L_\mathrm{iso}$ (\textit{solid lines}) and associated pressure bump locations (\textit{vertical dashed lines}). The initial vortex forms at $t = 60$ (\textit{black}). At $t = 1220$ (\textit{blue}), the outer gap edge is a ring and a new bump arises in $L_\mathrm{iso}$. At $t = 1440$ (\textit{light blue}), that bump causes a later-generation vortex to form. At $t = 2180$ (\textit{purple}), although another bump in $L_\mathrm{iso}$ arises, it is too far inside the pressure bump to form another vortex at the gap edge ($r_\mathrm{pressure} - r_\mathrm{crit} > 3.25~H_0$).} 
\label{fig:critical_examples}
\end{figure}

\begin{figure} 
\centering
\includegraphics[width=0.43\textwidth]{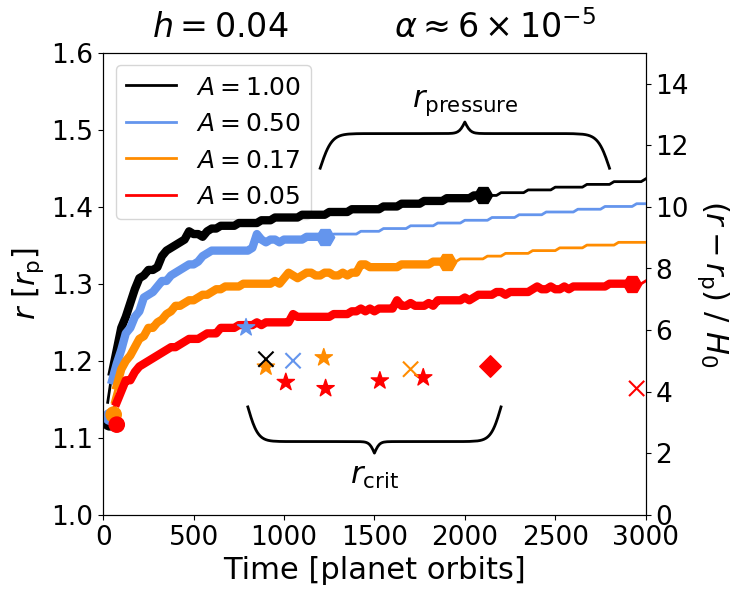} \\
\vspace{1em}
\includegraphics[width=0.43\textwidth]{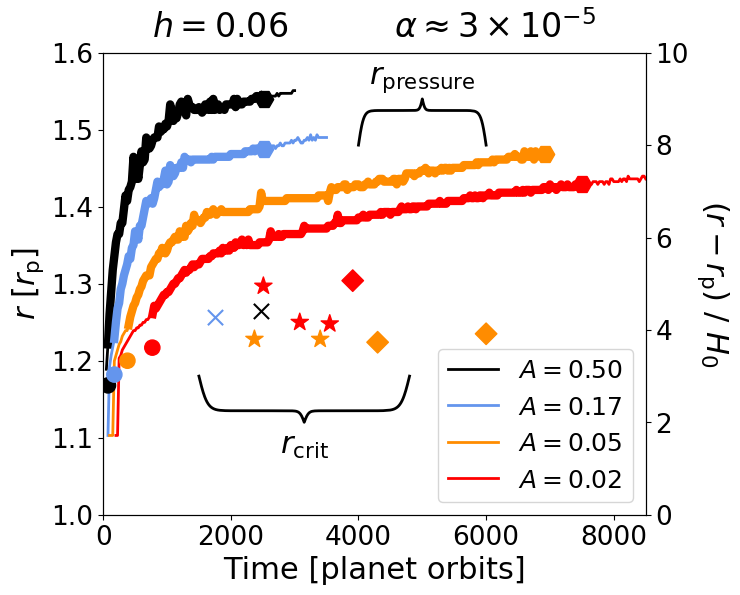} \\
\vspace{2em}
\includegraphics[width=0.38\textwidth]{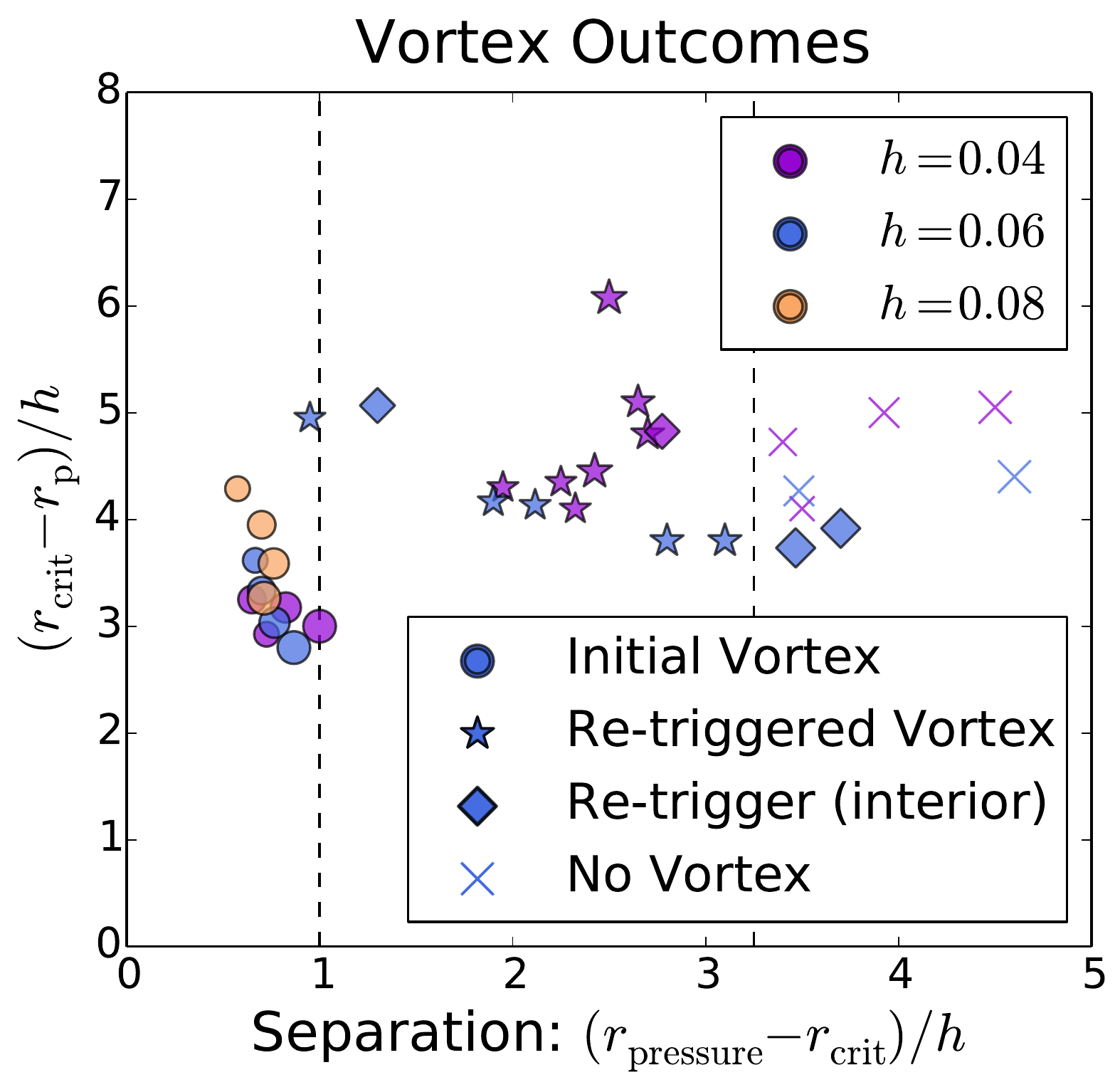} \hspace{1.5em}
\caption{\textit{Top and middle panels:} Locations of the pressure bump ($r_\mathrm{pressure}$) and the $L_\mathrm{iso}$ critical function maximum ($r_\mathrm{crit}$) as a function of time for cases with the lower viscosity $\nu = 10^{-7}$ and lower aspect ratios $h = 0.04$ and $h = 0.06$. A circle denotes the start of the vortex lifetime; a hexagon denotes the end. The lifetime itself is in bold. Each point for $r_\mathrm{crit}$ refers to a time at which a later-generation maximum arises in $L_\mathrm{iso}$. \textit{Bottom panel:} Different outcomes as a function of the separation between the two maxima. A new vortex forms if $r_\mathrm{pressure} - r_\mathrm{crit} < 3.25~H_0$. With larger separations, only interior vortices can form, if any form at all.} 
\label{fig:pressure_bumps}
\end{figure}

\begin{figure} 
\centering
\includegraphics[width=0.49\textwidth]{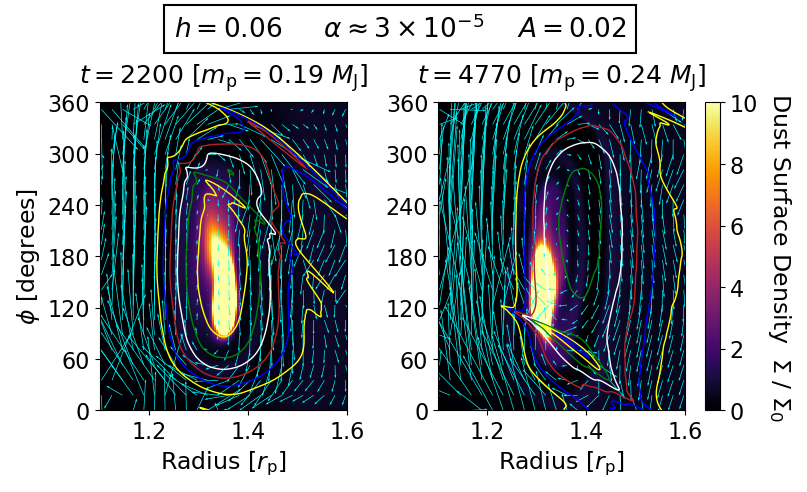} \\
\caption{Comparison of dust surface density in the initial vortex (\textit{left}) to an interior later-generation vortex (\textit{right}). Gas density contours {\cbf(at $\Sigma / \Sigma_0 = 0.8, 0.9, 1.0,$ etc.)} and gas velocity arrows are overlaid on both panels. Unlike the initial vortex, the interior re-triggered vortex is characterized by a much lower azimuthal gas density contrast and traps dust interior to the pressure bump ($r_\mathrm{pressure}$) closer to the critical function maximum ($r_\mathrm{crit}$). Regular re-triggered vortices (\textit{not shown}) resemble the initial vortex in having a high azimuthal density contrast and dust trapped at the location of the pressure bump.} 
\label{fig:interior}
\end{figure}

Planets that trigger short-lived initial gas vortices can still prolong the lifetimes of their associated dust asymmetry by re-triggering one or more later-generation vortices. When the initial vortex collapses and spreads into a ring in the $h = 0.04$ and $h = 0.06$ cases, it smooths out the the critical function so that it is too flat to become unstable again if it were left unchanged. Nonetheless, the profile is still subject to change because of the planet.

While no vortex is present, the planet can induce a drop in the vortensity and recreate a maximum in the critical function $L_\mathrm{iso}$ at the location with the peak negative vortensity change if the profile still has not reached a steady state. As shown in Figures~\ref{fig:critical_examples}~and~\ref{fig:pressure_bumps}, the maximum in the critical function and the pressure bump move further from the planet compared to when the outer gap edge initially becomes unstable to the RWI. More importantly, the separation between these two locations also increases. As a result, the new maximum in the critical function does not necessarily generate a later-generation vortex at the pressure bump.

The separation between these two locations results in three possible outcomes in the lower ($h = 0.04$) and intermediate ($h = 0.06$) aspect ratio cases:
\begin{enumerate}
  \item Re-triggered vortex: \\
  If the separation between the maximum in the critical function ($r_\mathrm{crit}$) and the pressure bump ($r_\mathrm{pressure}$) remains small ($r_\mathrm{pressure} - r_\mathrm{crit} < 3.25~H_0$), the disc becomes unstable at the location of the critical function maximum. The vortices that form at this location alter $L_\mathrm{iso}$ and shift the maximum much closer to the pressure bump. As a result, a new later-generation vortex typically forms at the pressure bump. Because the bump is only marginally unstable, this later-generation vortex starts out as an $m = 1$ mode and does not survive more than 100 orbits. 
  \item No vortex (with dust): \\
  If the separation between the maximum in the critical function and the pressure bump is too large ($r_\mathrm{pressure} - r_\mathrm{crit} > 3.25~H_0$), we do not find any cases where a vortex forms at the pressure bump. In lower-mass cases where $r_\mathrm{crit}$ and $r_\mathrm{pressure}$ are closer to the planet, an interior re-triggered vortex that is able to collect dust can still form (see the last case below). In higher-mass cases, it is also possible for vortices to form; however, they are so far interior to the pressure bump that the bump itself collects all of the dust drifting in from the outer disc and prevents any of it from reaching these vortices.
  \item Re-trigger (interior): \\
  In other cases, the later-generation vortex forms slightly interior to the pressure bump by about one scale height. Although this vortex is not at the center of the pressure bump, it is still close enough to collect the dust that accumulated at the pressure bump and trap it in an asymmetry interior to the bump (see Figure~\ref{fig:interior}). While these interior vortices do not create any substantial signature in the gas density structure, they can survive for several thousand orbits, much longer than the re-triggered vortices with a density signature at the bump itself and possibly longer than the initial vortex.
\end{enumerate}

The radial profiles of the critical function in the first and second cases are presented in Figure~\ref{fig:critical_examples} in comparison to the profile when the initial vortex formed. The times and locations of both maxima are shown in Figure~\ref{fig:pressure_bumps}. We find that $r_\mathrm{crit}$ typically stays at about 4 to $5~H_0$ away from the planet after the initial gas vortex dies and that this location is not strongly dependent on planet mass. On the other hand, $r_\mathrm{pressure}$ increases monotonically throughout each simulation due to the increasing planet masses and slow gap-opening timescales.

In the lower-mass cases, the first one or first few maxima in $L_\mathrm{iso}$ result in regular re-triggered vortices. These are followed by maxima that produce a mode of interior re-triggered vortices and then one or more ``no vortex" outcomes. In the higher-mass cases, the first maximum may already produce a ``no vortex" outcome because of their larger gap widths, skipping the re-triggered vortex phase, the interior re-triggered vortex, or both.

Regardless of whether a later-generation vortex forms, the ring that forms when the initial gas vortex dies is not initially perfectly axisymmetric. The disc maintains a remnant of the vortex in the form of an azimuthal pressure bump. Although it lacks vortical motion, this remnant pressure bump can sustain the existing dust asymmetry for a few hundred to a little over a thousand planet orbits before the dust spreads into a ring like the gas. As a result, even though the initial re-triggered vortices are very short-lived, they can greatly extend the total lifetime of the dust asymmetry, especially if more than one arise in succession. Additionally, the dust asymmetry from an early vortex always survives until the formation of the next-generation vortex. This behavior is the reason we define the dust asymmetry lifetime from the start of the first $m = 1$ vortex to the end of the last one.


\subsubsection{Role of spiral shocks} \label{sec:shocks}

\begin{figure} 
\centering
\includegraphics[width=0.47\textwidth]{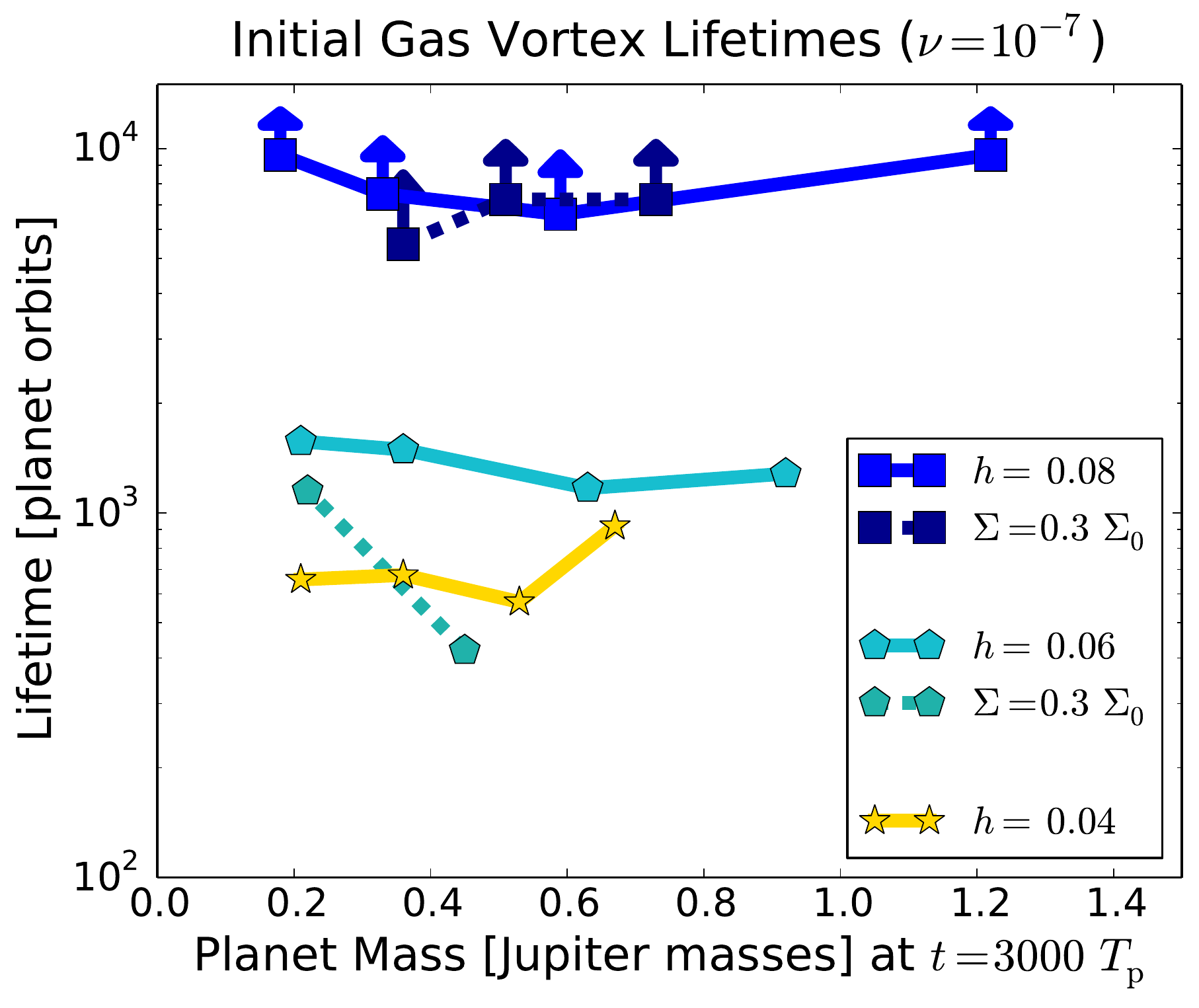}
\caption{Initial gas vortex lifetimes as a function of planet mass (recorded at $t = 3000$ planet orbits) for cases with a viscosity of $\nu = 10^{-7}$. Lower disc mass cases are denoted as ``$\Sigma = 0.3~\Sigma_0$". With $h = 0.08$, an asymmetry survives past the end of each simulation in all cases (as indicated by the arrows). Generally, the gas lifetime increases with increasing disc aspect ratio due to the planet's spiral shocks becoming weaker.} 
\label{fig:gas_lifetimes}
\end{figure}

Without taking the later-generation vortices into account, we find that the lifetimes of the initial gas vortices are primarily dependent on the disc aspect ratio and only weakly dependent on the planet mass (see Figure~\ref{fig:gas_lifetimes}). In our parameter study, all of the initial gas vortices with $h = 0.08$ live longer than those at $h = 0.06$, which in turn live longer than those at $h = 0.04$. In particular, the vortex in the second-highest mass case with $h = 0.06$ outlives the lowest-mass case with $h = 0.04$ even though both planets grow to roughly their respective values of the thermal mass. 

We suspect that trend results from the planet's spiral waves producing stronger shocks with lower disc aspect ratios. These shocks are stronger because the pre-shock Mach number is inversely proportional to the sound speed and in turn, the aspect ratio. We find that all cases have multiple shocks passing through the vortex, one from the planet itself and another from the material that collects at the $L_\mathrm{5}$ Lagrange point behind the planet. These spiral shocks play a strong role in disrupting the vortex and causing it to spread into a ring.

\section{In-depth Analysis (Focus: $\nu = 10^{-7}$)} \label{sec:gas-results}

In this section, we delve deeper into how the main results described in Section~\ref{sec:results} manifest in specific cases.\footnote{{\cbf With regard to the total dust asymmetry lifetimes discussed in this section that are shown in Figure~\ref{fig:lifetimes}, we note that we are tracking the degree of asymmetry in the dust, as discussed in Sections~\ref{sec:dust-simulations}~and~\ref{sec:signature}. On the other hand, the lifetimes of individual gas vortices are instead based on when the gas spreads into a ring or causes the appearance of the dust to change.}} We focus on the low-viscosity cases ($\nu = 10^{-7}$) throughout most of the section, exploring in depth both the intermediate aspect ratio cases with $h = 0.06$ in Section~\ref{sec:med-h} and the larger aspect ratio cases with $h = 0.08$ in Section~\ref{sec:large-h}. We then briefly highlight the key results for the lower aspect ratio cases with $h = 0.04$ in Section~\ref{sec:small-h} and all of the high-viscosity cases ($\nu = 10^{-6}$) in Section~\ref{sec:high-visc}.


\subsection{Intermediate Aspect Ratio ($h = 0.06$)} \label{sec:med-h}

\begin{figure*} 
\centering
\includegraphics[width=0.98\textwidth]{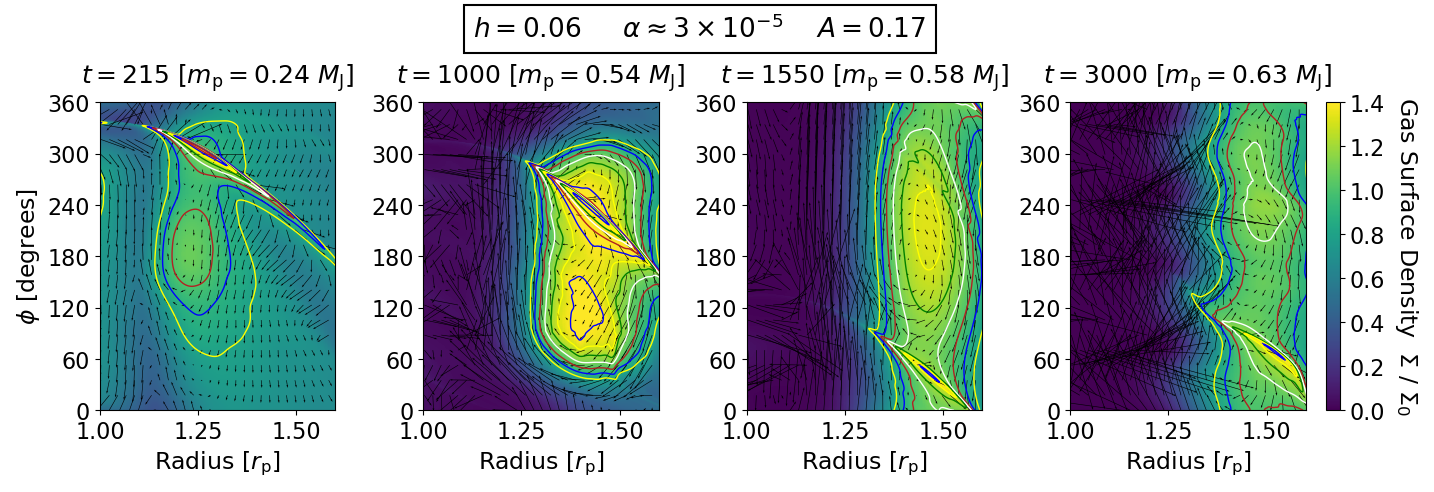} \\
\vspace*{0.5em}
\hspace*{1em}
\includegraphics[width=0.98\textwidth]{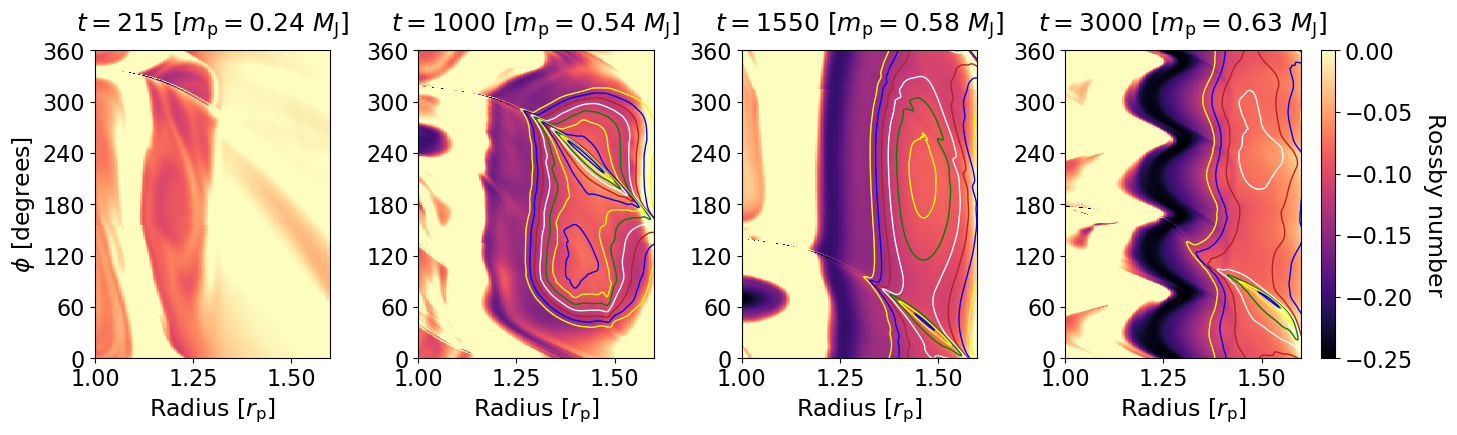}
\caption{Gas density (\textit{top panels}) and Rossby number evolution (\textit{bottom panels}) with a super-thermal mass planet ($h = 0.06$,  $\nu = 10^{-7}$, $A = 0.167$). Density contours {\cbf(at $\Sigma / \Sigma_0 = 0.8, 0.9, 1.0,$ etc.)} are overlaid on most panels. \textit{Column~1}: The $m = 2$ vortices centered at ($r / r_\mathrm{p}$, $\phi$) = ($1.20$, $180^{\circ}$) and ($1.25$, $300^{\circ}$) in the Rossby number map are divided by a diagonal line of positive vorticity that prevents them from merging into a compact vortex. \textit{Column 2}: The $m = 1$ elongated vortex{\cbf, in which the dust resembles the left panel of Figure~\ref{fig:interior}}. \textit{Column 3}: Although the vortex has spread out into a ring in the gas, the dust signature remains in the remnant of the pressure bump {\cbf and still resembles the left panel of Figure~\ref{fig:interior}.} \textit{Column~4}: Although a new bump in $L_\mathrm{iso}$ spawns dust-free vortices interior to the pressure bump, the bump itself is a ring in both the gas and dust.} 
\label{fig:evolution_h06_a167}
\end{figure*}

\begin{figure} 
\centering
\includegraphics[width=0.4\textwidth]{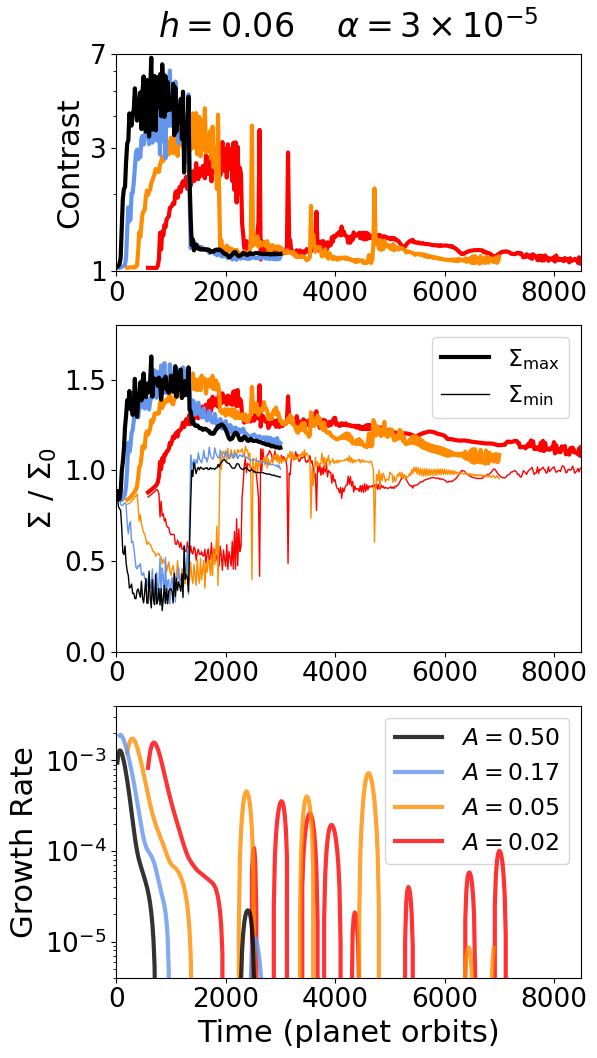}
\caption{\textit{Top:} Azimuthal density contrast ($\Sigma_\mathrm{max} / \Sigma_\mathrm{min}$). \textit{Middle:} $\Sigma_\mathrm{max}$ and $\Sigma_\mathrm{min}$ at location of pressure bump. \textit{Bottom:} Growth rate $d \ln \Delta \Sigma / dt$, where $\Delta \Sigma = \Sigma_\mathrm{max} - \Sigma_\mathrm{min}$. The growth of the RWI does not last 2000 orbits in any case. With $A = 0.02$ and 0.05, later-generation vortices form and are each identifiable by spikes in the contrast and the growth rate.} 
\label{fig:growth_h06}
\end{figure}

\begin{figure} 
\centering
\includegraphics[width=0.49\textwidth]{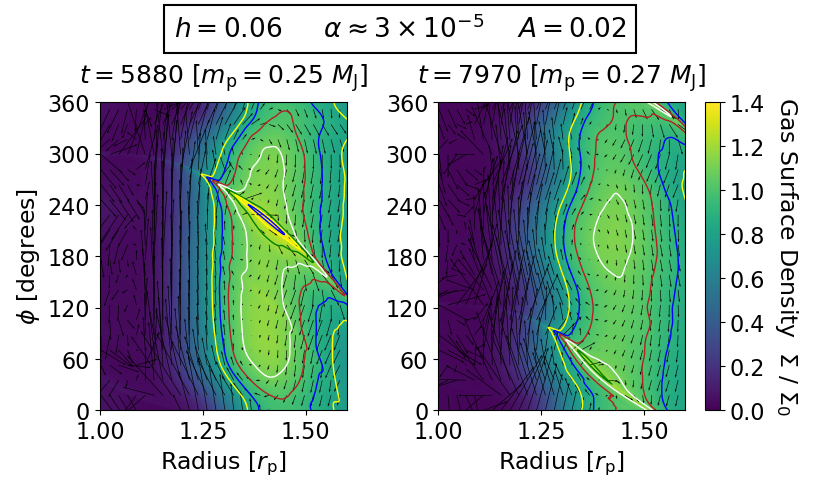} \\
\vspace*{0.5em}
\hspace*{0.5em}
\includegraphics[width=0.49\textwidth]{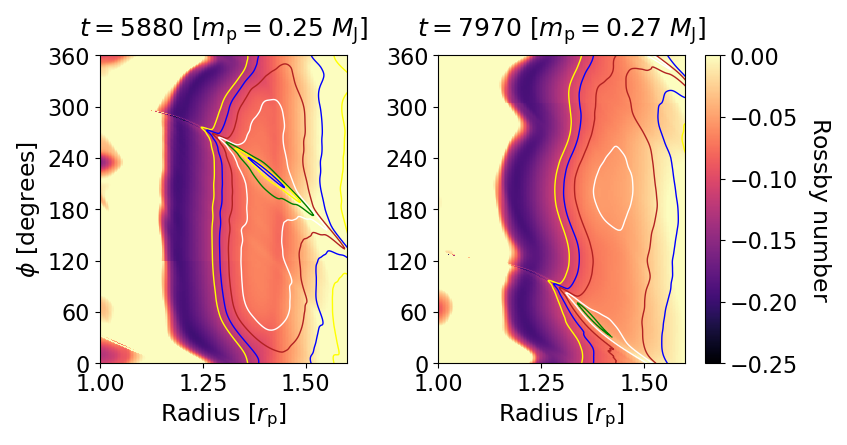}
\caption{Gas density (\textit{top panels}) and Rossby number late-stage evolution (\textit{bottom panels}) with a sub-thermal mass planet ($h = 0.06$,  $\nu = 10^{-7}$, $A = 0.02$). \textit{Column 1}: Interior later-generation vortex, which is long-lived. \textit{Column 2}: The interior vortex evolves into an $m = 3$ mode (see Figure~\ref{fig:dust_evolution_h06_a02}).} 
\label{fig:evolution_h06_a02}
\end{figure}

\begin{figure} 
\centering
\includegraphics[width=0.49\textwidth]{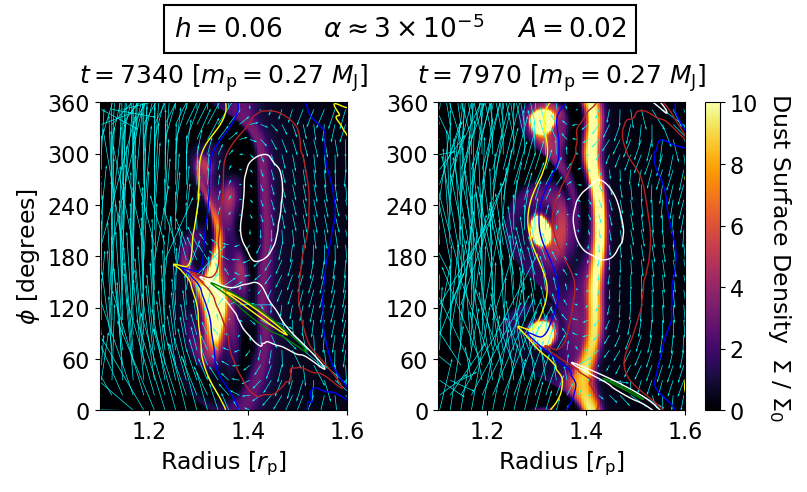} \\
\caption{Late-stage evolution of late-generation interior vortex. At $t = 7340$ (\textit{left}), the interior vortex is an $m = 1$ mode. By $t = 7970$ (\textit{right}), it changes to an $m = 3$ mode. In both snapshots, the vortices are accompanied by a ring of dust at the location of the pressure bump.}
\label{fig:dust_evolution_h06_a02}
\end{figure}

With an intermediate aspect ratio of $h = 0.06$, we simulate cases both above and below the thermal mass in our parameter study. The two higher-mass cases ($A = 0.5$ and 0.167) are near or above the thermal mass of $M_\mathrm{th} = 0.65~M_\mathrm{Jup}$, while the other two cases ($A = 0.05$ and 0.02) are well below the thermal mass. We discuss the general growth tracks of the planet in Appendix~\ref{sec:h06_planet}.

\subsubsection{Super-thermal mass} \label{sec:h06_super}

Both high-mass cases follow a similar evolution, as depicted in Figure~\ref{fig:evolution_h06_a167} by the $A = 0.167$ case that has $M_\mathrm{p,~3000} = 0.63~M_\mathrm{Jup}$. Once the outer gap edge becomes unstable, it takes 40 to 50 orbits for the initial $m = 3$ mode of the RWI to saturate and merge into a single $m = 1$ elongated vortex that lacks smooth elliptical contours because of the interaction with the planet's two exterior spiral waves. Even after this vortex materializes, the instability continues to grow non-linearly at an extremely slow rate for about 1000 orbits, as illustrated by the evolution of the vortex's density contrast in Figure~\ref{fig:growth_h06}. Within a few hundred orbits after the growth of the instability stops, the vortex suddenly collapses into a ring over just a few orbits.

When a new maximum in $L_\mathrm{iso}$ arises afterwards, it is located too far interior to the bump to generate any signature in the dust. Nonetheless, the dust asymmetry associated with the remnant azimuthal pressure bump persists for an additional 1000 orbits after the initial gas vortex collapsed into a ring. As a result, the total lifetime of the dust asymmetry is $\approx 2400$ orbits in both cases (see the aqua lines with pentagon points in Figure~\ref{fig:lifetimes}).

We also tested whether this behavior is independent of disc mass with an additional simulation that has $70\%$ lower disc mass and a planet that grows to $M_\mathrm{p,~3000} = 0.46~M_\mathrm{Jup}$ with $A = 1.0$. Although we found a similar total lifetime of 2100 orbits, the initial vortex only survives about 500 orbits, likely due to the deeper gap resulting in the vortex not developing as high of a contrast as in those other cases. With this shorter lifetime, however, the planet can generate three later-generation vortices because the pressure bump is still relatively close to the planet early in the simulation. By the time a fourth maximum arises though, $r_\mathrm{pressure} - r_\mathrm{crit} > 3.25~H_0$, there is too big of a separation to create another vortex. This test confirms that higher-mass planets still generate shorter-lived dust asymmetries with lower disc masses and also demonstrates that the boundary between regimes is not exactly the thermal mass.

\subsubsection{Sub-thermal mass} \label{sec:h06_sub}

Both low-mass cases follow a similar evolution to each other. Although the initial evolution is also similar to the high-mass cases (aside from starting out with an $m = 1$ mode in the lowest-mass case), the location of the pressure bump remains relatively close to the $L_\mathrm{iso}$ maximum ($r_\mathrm{pressure} - r_\mathrm{crit} < 3.25~H_0$) after the initial gas vortex collapses in about 1500 orbits. As a result, both cases produce two short-lived regular later-generation vortices. Afterwards, the third maxima that arise in $L_\mathrm{iso}$ each yield a long-lived interior vortex that survives for over 2000 orbits. As a result, the total lifetimes of both $m = 1$ dust asymmetries are $\approx 6000$ orbits, about 2.5 times longer than the vortices in the super-thermal mass regime.

Aside from the usual $m = 1$ dust asymmetry, both low-mass cases can also generate an $m \ge 3$ dust asymmetry after the initial $m = 1$ vortex collapses. 
\begin{enumerate} 
  \item In particular in the lowest-mass case ($A = 0.02$ case that has $M_\mathrm{p,~3000} = 0.21~M_\mathrm{Jup}$), when the $m = 1$ interior dust asymmetry disappears at $t = 7500~T_\mathrm{p}$, it does not spread into a ring. Instead, the mode of the RWI changes from $m = 1$ to $m = 3$ (see Figure~\ref{fig:evolution_h06_a02}), resulting in three dust-trapping vortices at the same separation as the previous interior one (see Figure~\ref{fig:dust_evolution_h06_a02}). Like the $m = 1$ interior vortices, these $m = 3$ vortices do not create a noticeable perturbation in the density and are much easier to identify in the vortical motion of the velocity field or using dust to trace them out. These $m = 3$ vortices survive at least another 1500 orbits, at which point we stop the simulation because they do not show any sign of decaying. 
  \item Similarly in the second-lowest mass case ($A = 0.02$ case that has $M_\mathrm{p,~3000} = 0.36~M_\mathrm{Jup}$), an $m = 4$ set of vortices arises at $t = 3500~T_\mathrm{p}$ while only a remnant pressure bump was present. Unlike the lowest-mass case, these vortices only survive about 50 orbits, at which point they alter the $L_\mathrm{iso}$ function and generate a regular short-lived later-generation vortex at the pressure bump.
  \end{enumerate}
We suspect these interior $m \ge 3$ vortices can survive due to the large gradient in $L_\mathrm{iso}$ at the interior side of the outer gap edge pressure bump. In general, Rossby waves arise at sharp gradients in the vortensity or $L_\mathrm{iso}$ \citep{umurhan10}. With strong Rossby waves present at the location of these vortices, these sharp gradients can act to sustain the vortices generated by the instability. 

Like the super-thermal test with a lower disc mass, we found that the general sub-thermal behavior does not depend on disc mass. We ran an additional simulation with a $70\%$ lower disc mass and a planet that grows to $M_\mathrm{p,~3000} = 0.24~M_\mathrm{Jup}$ as $A = 0.125$. Like with our fiducial disc mass, the planet generates a series of later-generation vortices, the last of which is an interior vortex that forms at $t = 3900~T_\mathrm{p}$ when the separation between the maxima is $r_\mathrm{pressure} - r_\mathrm{crit} < 3.25~H_0$. As a result, the asymmetry survives for $4700~T_\mathrm{p}$ in total, nearly as long as in both sub-thermal cases from our parameter study. Overall, the two lower disc mass tests verify that the planet mass is more important than the disc mass in determining the behavior of the vortices in the low aspect ratio cases. 



\subsubsection{Role of spiral waves (with lower fixed planet mass)} \label{sec:h06_lower}

Since none of the initial gas vortices in our parameter study are particularly long-lived with $h = 0.06$, we ran an additional simulation with an even lower fixed mass of $M_p~=~0.08~M_\mathrm{Jup}$. We found that the initial vortex in this case survives at least 4400 orbits from $t = 600$ to $>5000~T_\mathrm{p}$, at which point we stop the simulation because the vortex does not show any sign that it is close to collapsing into a ring. As indicated by the azimuthal density contrast, the instability continues to grow up until $t \approx 2000~T_\mathrm{p}$ when the contrast reaches a peak value of $\approx 1.5$ before decaying at a very slow rate for the rest of the simulation.

Unlike in our parameter study where the vortex is repeatedly disrupted by two exterior waves for nearly its entire lifetime, the vortex in this case is initially only disrupted by the one spiral wave directly induced by the planet. The second wave does not appear until $t \approx 3000~T_\mathrm{p}$ when the planet finally clears out enough of the material in the gap for the remaining material to collect at the $L_5$ Lagrange point. Moreover, both waves are associated with weaker density perturbations because the planet has a lower mass. As a result, the initial vortex never abruptly collapses and ultimately survives for more than three times as long as the $h = 0.06$ initial vortices from our parameter study. We observe similar behavior in the low disc mass cases with $h = 0.08$ that also have weaker spiral shocks (see Section~\ref{sec:h08_disc}).

We further tested the role of the planet's spiral waves by weakening the shocks in the $A = 0.02$ case. To do that, we restarted the simulation at $t = 810~T_\mathrm{p}$ when the planet has a mass of $0.11~M_\mathrm{Jup}$ and did not allow the planet to accrete any more mass. As a result, the second wave from the $L_5$ Lagrange point takes longer to emerge, the waves do not strengthen over time, and the initial vortex survives an additional 350 orbits compared to in our parameter study. This longer lifetime verifies that the strength of the spiral shocks primarily determines if and when the vortices spread into rings in the $\nu = 10^{-7}$ cases.

\subsubsection{Role of viscosity} \label{sec:h06_collapse}

We demonstrated that the planet's spiral waves are responsible for causing the initial vortex to collapse with the fixed lower-mass case. Does the viscosity also play a role in killing the vortex? We tested this possibility with the $A = 0.167$ case where the gas vortex spreads into a ring at $t = 1347~T_\mathrm{p}$ when the planet has a mass of $0.57~M_\mathrm{Jup}$. More specifically, we restarted the simulation at $t = 1000~T_\mathrm{p}$ with the prescribed viscosity reduced to zero.

We found that the vortex survives until at least the end of the simulation at $t = 3000~T_\mathrm{p}$, well beyond the end of the lifetime in our parameter study. We find that with no prescribed viscosity, the vortex and the instability continue to grow at a rather slow rate as indicated by the evolution of the density contrast. This behavior demonstrates that the non-zero viscosity damps and eventually halts the growth of the instability in our parameter study, which in turn leads to the spiral waves causing the vortex collapse shortly after the instability stops growing.

\subsection{Larger Aspect Ratio ($h = 0.08$)} \label{sec:large-h}

\begin{figure} 
\centering
\includegraphics[width=0.4\textwidth]{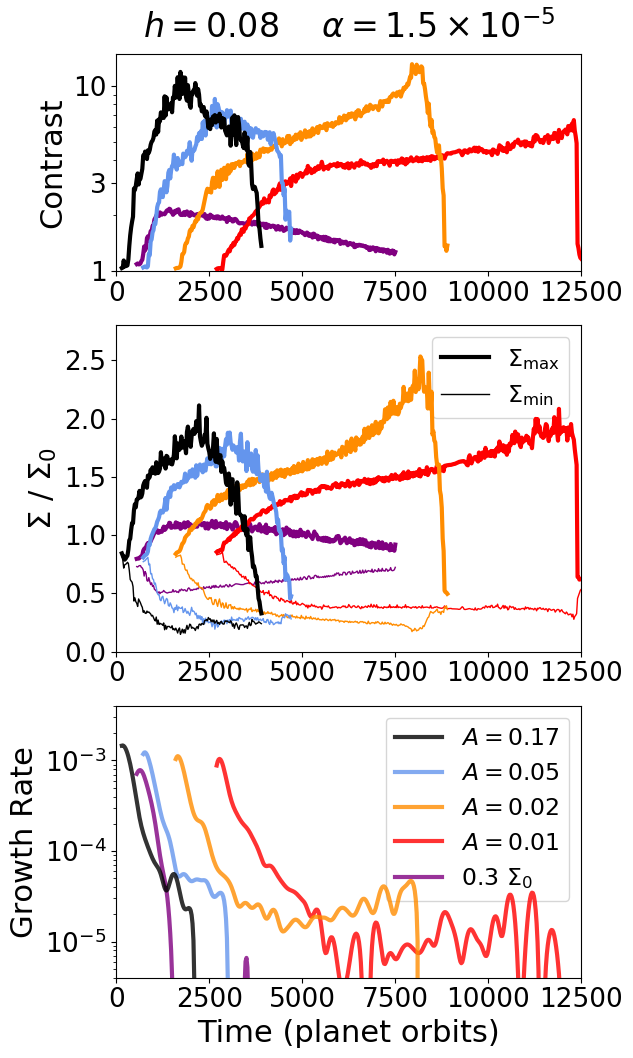}
\caption{\textit{Top:} Azimuthal density contrast ($\Sigma_\mathrm{max} / \Sigma_\mathrm{min}$). \textit{Middle:} $\Sigma_\mathrm{max}$ and $\Sigma_\mathrm{min}$ at location of pressure bump. \textit{Bottom:} Growth rate $d \ln \Delta \Sigma / dt$, where $\Delta \Sigma = \Sigma_\mathrm{max} - \Sigma_\mathrm{min}$. In our parameter study, the RWI grows until the vortex becomes compact and destroys itself through its spiral waves. With a low disc mass ($0.3~\Sigma_\mathrm{0}$), although the RWI stops growing early, the vortex survives longer than the higher-mass cases from our parameter study.} 
\label{fig:growth_h08}
\end{figure}

\begin{figure*} 
\centering
\includegraphics[width=0.42\textwidth]{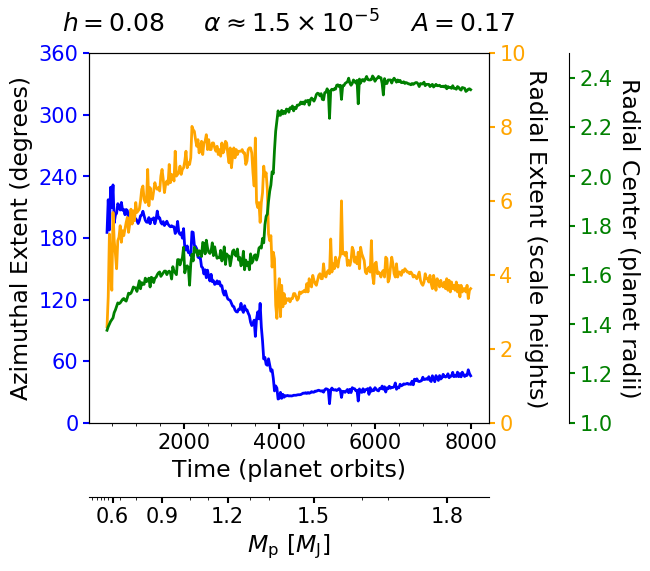}\hspace{2em}
\includegraphics[width=0.42\textwidth]{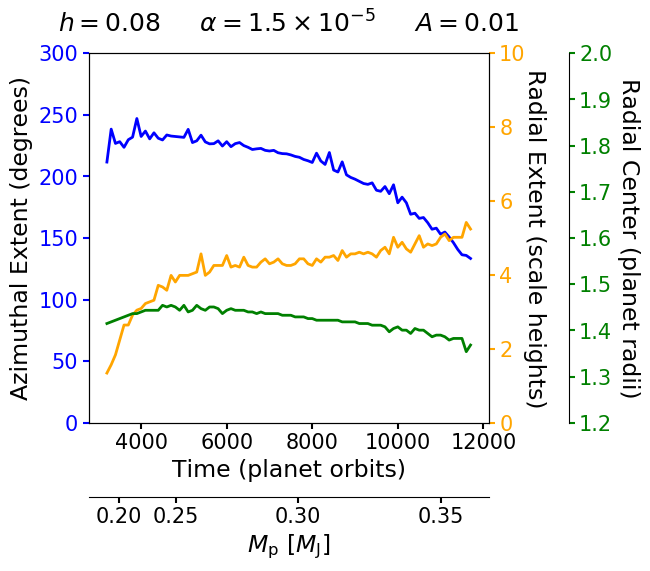}
\caption{Vortex azimuthal extent (blue), radial extent (orange), and radial center (green) as a function of time for cases with $h = 0.08$ and $\nu = 10^{-7}$ ($\alpha \approx 1.5 \times 10^{-5}$). In all four cases, the vortex becomes more compact (blue) and becomes wider (orange) over time during its lifetime. The vortices in the two higher-mass cases \cbf{(e.g. \textit{left panel})} migrate outwards, while the vortices in the two lower-mass cases \cbf{(e.g. \textit{right panel})} migrate inwards.} 
\label{fig:extents_h08}
\end{figure*}

\begin{figure*} 
\centering
\includegraphics[width=0.98\textwidth]{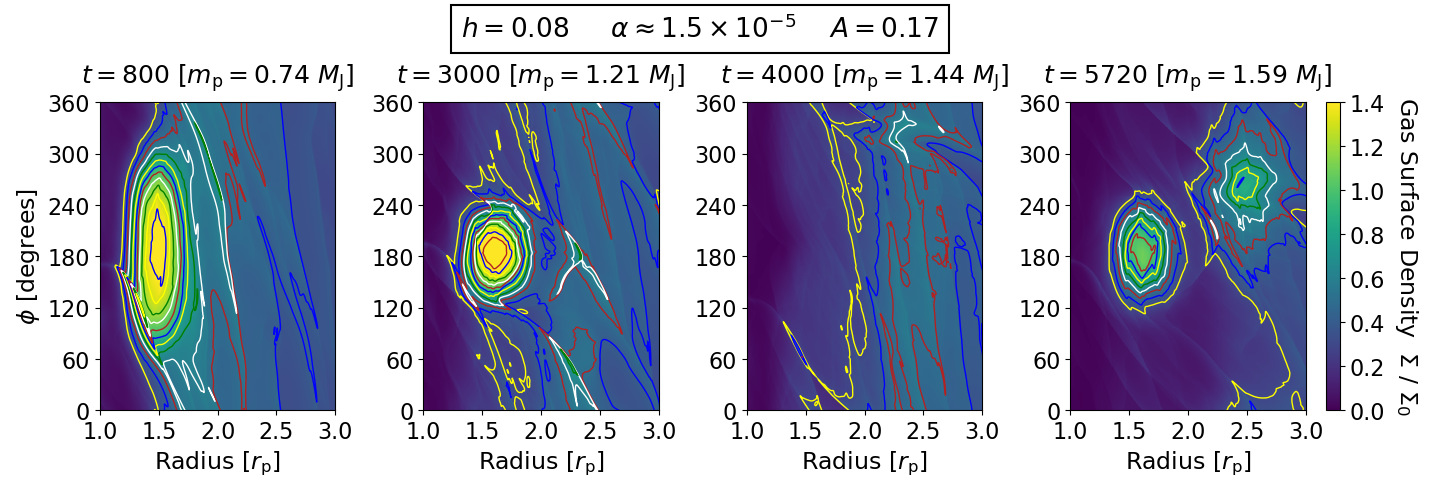} \\
\vspace*{0.5em}
\hspace*{1em}
\includegraphics[width=0.98\textwidth]{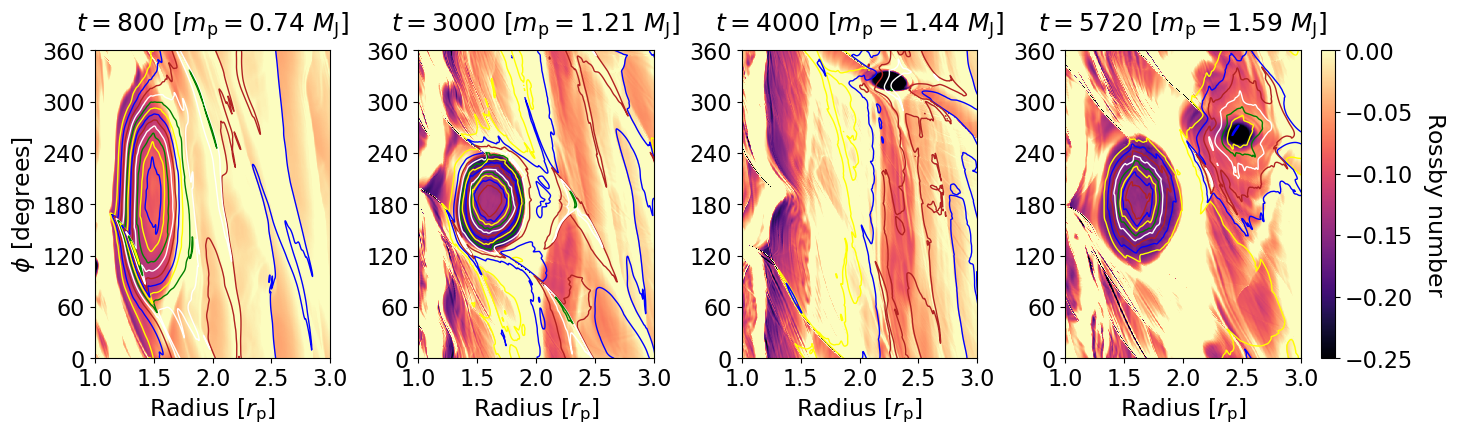}
\caption{Gas density (\textit{top panels}) and Rossby number evolution (\textit{bottom panels}) in the highest-mass case ($A = 0.167$) with $h = 0.08$ and $\nu = 10^{-7}$. Density contours {\cbf(at $\Sigma / \Sigma_0 = 0.3, 0.4, 0.5,$ etc.)} are overlaid on all panels. \textit{Column~1}: The initial elongated vortex. \textit{Column 2}: The vortex has transitioned to compact ($\mathrm{Ro} < -0.15$) and is still shrinking.  \textit{Column 3}: The initial vortex has migrated to $r > 2.0~r_\mathrm{p}$ and a new elongated vortex has formed in its place at $r \sim 1.4~r_\mathrm{p}$. \textit{Column~4}: The later-generation vortex has become more compact.} 
\label{fig:evolution_h08_a167}
\end{figure*}

Just like in discs with lower aspect ratios, we find that slower-grown planets with lower masses also trigger longer-lived vortices in discs with a large aspect ratio of $h = 0.08$, albeit through two separate pathways. Each case in our parameter study features a planet below the thermal mass ($M_\mathrm{th} = 1.54~M_\mathrm{Jup}$). In all four cases, the vortex is sustained by the RWI continuing to grow for thousands of orbits, as emphasized in Figure~\ref{fig:growth_h08}. Eventually, the vortex becomes strong enough for its own waves to drastically shrink the vortex by dissipating a significant amount of angular momentum (see Figure~\ref{fig:extents_h08}). With this dissipation, the evolution of the RWI in the outer disc is more complicated than any of the other cases, as illustrated in Figure~\ref{fig:evolution_h08_a167}. On the other hand in lower-mass discs, the planet opens up a deeper gap as a result of accreting a much larger fraction of the gas in its vicinity. Although that depletion halts the growth of the instability, these vortices are still long-lived because the waves that shock the vortex are weaker with a larger disc aspect ratio.

Unlike in discs with lower aspect ratios, the RWI can continue to grow at a high rate for thousands of orbits because the shallow gap depths allow the planet to keep growing at an appreciable rate over that period of time (as shown in Figure~\ref{fig:growth_h08}). The instability grows unimpeded until the following critical points in each case:
\begin{itemize}[leftmargin=1em]
\item ~[$A = 0.17$]:~~~$t \approx 2000~T_\mathrm{p}$~~~($m_\mathrm{p} \approx 0.97~M_\mathrm{J}$),~~~$\delta_\mathrm{gap} \approx 42.7$
\item ~[$A = 0.05$]:~~~$t \approx 2750~T_\mathrm{p}$~~~($m_\mathrm{p} \approx 0.57~M_\mathrm{J}$),~~~$\delta_\mathrm{gap} \approx 12.6$
\item ~[$A = 0.02$]:~~~$t \approx 8150~T_\mathrm{p}$~~~($m_\mathrm{p} \approx 0.48~M_\mathrm{J}$),~~~$\delta_\mathrm{gap} \approx 10.8$
\item ~[$A = 0.01$]:~~~$t \approx 11700~T_\mathrm{p}$~~~($m_\mathrm{p} \approx 0.36~M_\mathrm{J}$),~~~$\delta_\mathrm{gap} \approx 8.8$.
\end{itemize}
The three lower-mass gaps maintain shallow gap depths $\delta_\mathrm{gap} < 20$ up until those times. Although the highest-mass planet creates a deeper gap by about 1000 orbits, it still continues to grow significantly up until it reaches that critical point. Unlike the lower aspect ratio cases, viscosity never stops the instability from growing.


\begin{figure} 
\centering
\includegraphics[width=0.4\textwidth]{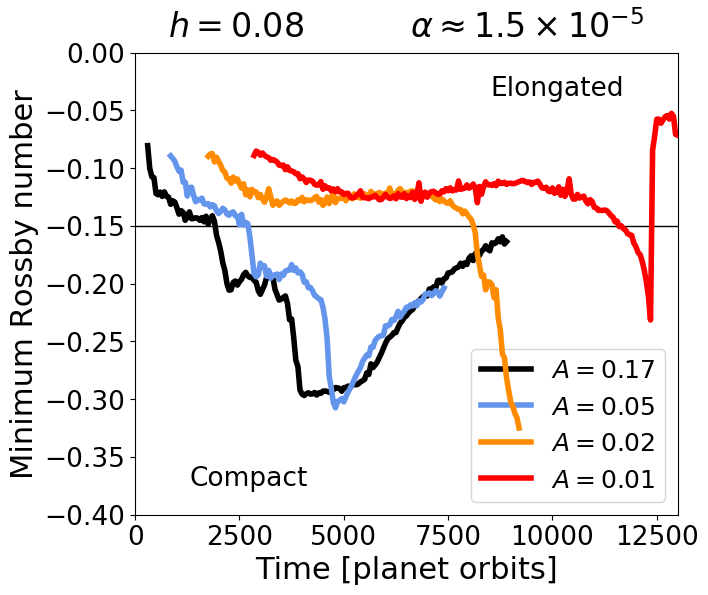}
\caption{Evolution of minimum Rossby number in vortices with $h = 0.08$ and $\nu = 10^{-7}$ ($\alpha \approx 1.5 \times 10^{-5}$). When the minimum Rossby number falls below $\mathrm{Ro} = -0.15$, it drops significantly and the vortex transitions from elongated to compact. The Rossby number undergoes a second major drop when the vortex migrates outward in the three higher-mass cases ($A \ge 0.02$). With such a low Rossby number, these vortices survive past the ends of the simulations.} 
\label{fig:rossby_number_h08}
\end{figure}

Instead, the vortex eventually develops waves that are strong enough to halt the growth of the instability. As the vortex's peak density and contrast both gradually increase over time, the vortex that was initially elongated also slowly develops a clear vorticity minimum at its center. When this minimum reaches a critical value of $\mathrm{Ro} = -0.15$ at the critical times listed above, the vortex's general structure changes from elongated to compact (see Figure~\ref{fig:rossby_number_h08}). As \cite{surville15} have demonstrated, a vortex with a compact structure and a pronounced vorticity minimum also launches much more pronounced spiral waves compared to when the vortex was elongated. Just like with a planet, the vortex launches two waves: one that leads the vortex out of the front azimuthal side and spirals inwards, and another that trails the vortex out of the back and spirals outwards.

The lack of symmetry between the inner and outer waves enables the vortex to transport angular momentum through the disc, which in turn typically causes the vortex itself to lose angular momentum. As \cite{paardekooper10} derived, the angular momentum excess of a vortex is given by
\begin{align} \label{shrink}
\mathcal{J} & \simeq \int^{r_0 + s}_{r_0 - s} dr r^2 (\delta \Sigma r \Omega(r) + \Sigma u_{\phi}) \Delta \phi \\
& \sim \Sigma_0 \omega_\mathrm{rot} s^3 r_0 \Delta \phi [- \frac{r_0^2 \Omega(r_0)^2}{c_s^2} + 1],
\end{align}
where $\Delta \phi$ is the azimuthal extent of the vortex, $s$ is the radial half-width of the vortex, $r_0$ is the radial center of the vortex, $\Sigma_0$ is the unperturbed density without the vortex, $\delta \Sigma$ is the perturbation of the density due to the vortex, $u_\mathrm{\phi}$ is the relative azimuthal velocity in the vortex, and $\omega_\mathrm{rot}$ is the rotation rate of the vortex. It is also assumed that the vortex is in geostrophic balance, that is, $\Omega(r_0) \omega_\mathrm{rot} s^2 \sim - c_s^2 \delta \Sigma / \Sigma_0$. Although the vortex is anticyclonic ($\omega_\mathrm{rot} < 0$), the vortex represents a positive excess of angular momentum ($\mathcal{J} > 0$) due to the large density perturbation associated with the vortex (i.e. the magnitude of the first term is larger than the second in Equation~\ref{shrink}). In order for the vortex to lose angular momentum, it needs to either migrate inwards (lower $r_0$), reduce its radial extent (lower $s$), or reduce its azimuthal extent (lower $\Delta \phi$). 

We find that the vortex preferentially loses angular momentum by shrinking in the azimuthal direction. In a smooth disc with a standard negative density gradient, vortices typically shed angular momentum by migrating inwards \citep{paardekooper10}. On the contrary, the vortices in our simulations are located at pressure bumps that can stop them from migrating inwards. Moreover, these vortices typically migrate outwards as the planet gradually pushes the pressure bump at the outer gap edge away. As a result, the vortex changes its shape instead. We suspect the vortex specifically shrinks in azimuth because vortices with vorticity minima at their centers are typically associated with lower, more compact aspect ratios \citep{surville15}. Additionally, as the vortex shrinks and transports angular momentum outwards, it funnels gas in its vicinity towards the star, filling some of the gap and increasing the accretion rate onto the planet.

\subsubsection{Excess accretion onto the planet} \label{sec:h08_accretion}

The planets in the $h = 0.08$ cases have irregular growth tracks because of the supplemental accretion due to the vortex's stronger spiral waves. Early in the vortex lifetime, the accretion rate from the vortex is negligible and the total accretion rate follows a predictable declining pattern based on the mass of the planet and the depth of the gap (see Figure~\ref{fig:mass}). At some point though, the accretion rates in all four cases eventually reach a minimum and then begin to increase, indicating that the vortex-induced accretion rate begins to overtake the standard viscous accretion rate.  These minima are reached at 
\begin{itemize}[leftmargin=1em]
\item ~[$A = 0.17$]:~~~$t \approx 1600~T_\mathrm{p}$~~~($m_\mathrm{p} \approx 0.90~M_\mathrm{J}$),~~~$\delta_\mathrm{gap} \approx 37.9$
\item ~[$A = 0.05$]:~~~$t \approx 2200~T_\mathrm{p}$~~~($m_\mathrm{p} \approx 0.52~M_\mathrm{J}$),~~~$\delta_\mathrm{gap} \approx 14.7$
\item ~[$A = 0.02$]:~~~$t \approx 5000~T_\mathrm{p}$~~~($m_\mathrm{p} \approx 0.39~M_\mathrm{J}$),~~~$\delta_\mathrm{gap} \approx 14.5$
\item ~[$A = 0.01$]:~~~$t \approx 7800~T_\mathrm{p}$~~~($m_\mathrm{p} \approx 0.30~M_\mathrm{J}$),~~~$\delta_\mathrm{gap} \approx 10.1$,
\end{itemize}
times which also correspond to local maxima in the gap depth due to the increased accretion rates filling the gap. All of the minima occur before the vortex reaches the critical point at $\mathrm{Ro} = -0.15$, showing that the waves can transport angular momentum before the vortex develops a compact structure and starts to rapidly shrink in azimuth. The minima in the two higher-mass cases precede the critical times by only a few hundred orbits, while the accretion rates in the other two cases start to increase thousands of orbits before the critical times. This added accretion due to the waves is responsible for keeping the gap shallow. It also provides a means for the planet to continue accreting at a high rate even after it has opened a gap, something that does not happen otherwise due to the disc's low viscosity.


\subsubsection{Late-stage evolution of initial vortex} \label{sec:h08_late}

\begin{figure} 
\centering
\includegraphics[width=0.42\textwidth]{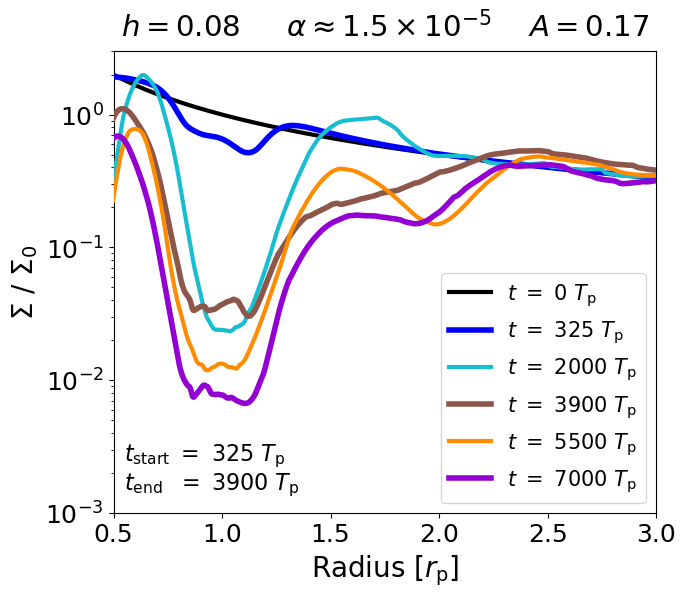}
\caption{Radial gas density (azimuthally-averaged) profiles over time for the highest-mass case with $h = 0.08$ and $\nu = 10^{-7}$ ($\alpha \approx 1.5 \times 10^{-5}$). The initial vortex forms at $t = 325$, reaches peak contrast and starts to become compact at $t = 2000$. It migrates outwards at $t = 3900$ due to the positive density gradient, and a second vortex forms in its place. This later-generation vortex reaches peak contrast and starts to become compact at $t = 5500$ before disappearing completely at $t = 7000$. The decay of both vortices keeps the gap shallow, allowing the planet to accrete a substantial amount of mass up until that point.} 
\label{fig:avg_density_h08}
\end{figure}

Once the critical time is reached, the rate at which the vortex becomes more compact accelerates. In the two higher-mass cases, the RWI still continues to grow for a few hundred orbits until the density starts to decline rapidly. In the other two cases, that density decline begins almost immediately. Meanwhile, the Rossby number minimum also drops (see Figure~\ref{fig:rossby_number_h08}). It reaches a higher amplitude value that the vortex maintains until it depletes enough of the density in its vicinity to remove the initial pressure bump {\cbf (see Figure~\ref{fig:avg_density_h08})}. The flow of that density into the gap leaves the vortex in a region of the disc with a positive density gradient. As \cite{paardekooper10} found, that positive density gradient causes the vortex to migrate outwards until it reaches the new pressure bump located at $r \ge 2.0~r_\mathrm{p}$. While the vortex is migrating, its vorticity minimum continues to drop. This higher magnitude vorticity allows the vortices to maintain their strong spiral waves and migrate outward even though their density signatures almost completely vanish.

After the vortex reaches the new outer pressure bump, it continues to survive for at least several thousand more orbits. Not long after the vortex has migrated, however, a new later-generation vortex forms at the location of the initial pressure bump, thereby leaving the disc with a different signature than the original $m = 1$ dust asymmetry. In the highest-mass case, the original vortex merges with a low-amplitude elongated $m = 1$ vortex at the outer pressure bump. Over time, this vortex slowly migrates inwards with the outer pressure bump and the amplitude of its vorticity gradually drops towards zero. Due to the much more negative minimum Rossby number at its center, the vortex does not decay very quickly. Ultimately, all of the initial vortices survive until at least 10000 orbits, and each of the initial vortices in the three highest-mass cases are still present at the end of their respective simulations.

The lone exception to the pattern of outward migration is the lowest-mass case, in which the vortex instead continues to migrate inwards. The vortex in this case had already been migrating inwards before it started to become more compact. Eventually, the inner edge of the vortex becomes co-orbital with the planet, at which point the vortex quickly decays completely even though the amplitude of the minimum Rossby number at its center had increased to a much higher value. We suspect the vortex dissipates due to its interactions with the planet.

\subsubsection{Later-generation vortices} \label{sec:h08_later}

\begin{figure} 
\centering
\includegraphics[width=0.43\textwidth]{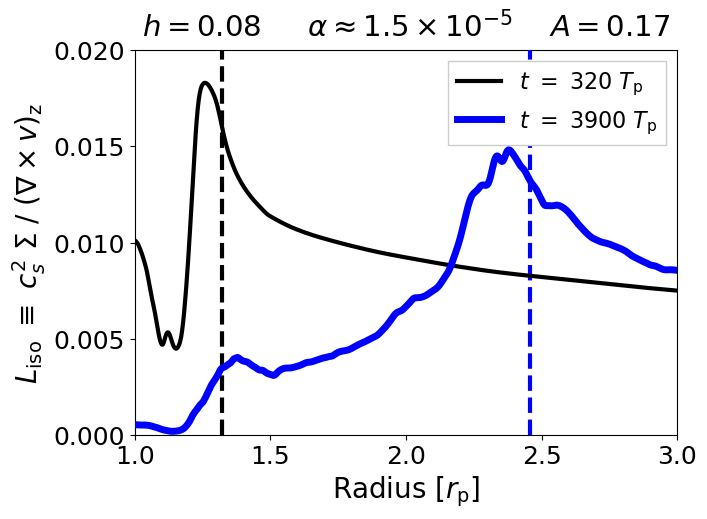} \\
\caption{Radial profiles of the RWI critical function $L_\mathrm{iso}$ (\textit{solid lines}) and associated pressure bump locations (\textit{vertical dashed lines}) for $h = 0.08$ and $A = 0.167$. The initial vortex forms at $t = 60$ (\textit{black}). By $t = 3900$ (\textit{blue}), the initial vortex has migrated outward, allowing a new later-generation vortex to form at the new bump in $L_\mathrm{iso}$ near $r = 1.4~r_\mathrm{p}$.} 
\label{fig:later-generation_h08}
\end{figure}

The outward migration of the initial vortices clears out space for a later-generation vortex to form in its place. With the initial vortex no longer competing against the planet to fill the gap, the planet resumes opening the gap and creates a new maximum in the critical function $L_\mathrm{iso}$ at the location of the peak negative vortensity change, as shown in Figure~\ref{fig:later-generation_h08}. These two factors are essentially the same as the two factors responsible for the generation of the initial vortex. The new maxima arise at 1.20 to $1.25~r_\mathrm{p}$, closer in than the maxima responsible for the initial vortices. These new maxima become unstable and generate $m = 1$ elongated vortices during the phase when the initial vortices are rapidly migrating outwards.

Like the initial vortices, these later-generation vortices are also relatively long-lived and follow a similar evolutionary pathway. These vortices start out elongated and then continue to grow long after they formed. As the vortex grows, it migrates outwards slightly. Eventually the vortex grows enough to establish a new pressure bump located at $r \approx 1.6~r_\mathrm{p}$, roughly the same location as the initial pressure bump. Once this pressure bump forms, the vortex stops migrating. At its new location, the vortex also grows enough to transition from elongated to compact, at which point it starts to become more compact due to the emission of its own spiral waves. Unlike the initial vortices, they cannot migrate outward to the exterior pressure bump during this phase because that is where the initial vortex is still located. Instead, they continue to shrink until they fade completely. We suspect this vortex completely disappears as a result of interacting with the stronger initial vortex. The later-generation vortex in the highest-mass case survives for 3100 orbits (until $t = 7000~T_\mathrm{p}$), while the corresponding vortex in the second highest-mass case survives for at least 2700 orbits (past the end of the simulation).

In the lowest-mass case, a later-generation vortex also forms because of the same two factors. With no other vortex present in the outer disc, however, we expect this vortex to be longer-lived than the other later-generation vortices, albeit we did not track its evolution beyond a few hundred orbits.

\subsubsection{Role of spiral waves (with lower disc mass)} \label{sec:h08_disc}

\begin{figure} 
\centering
\includegraphics[width=0.49\textwidth]{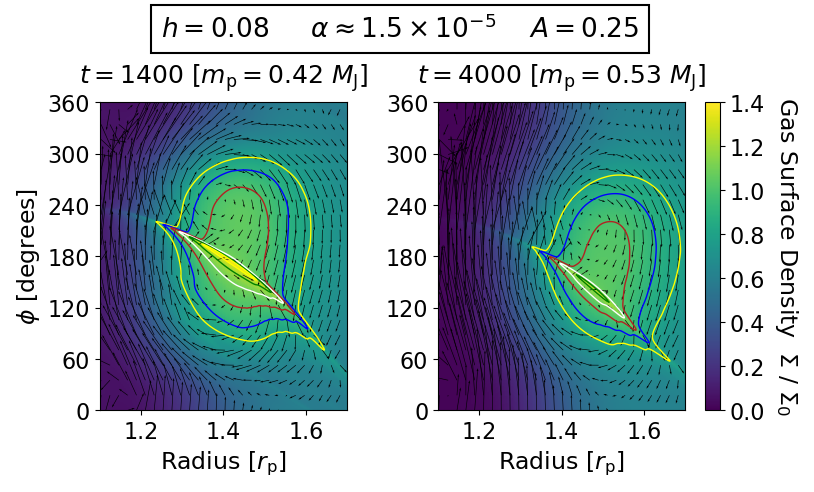} \\
\vspace*{0.5em}
\hspace*{0.5em}
\includegraphics[width=0.49\textwidth]{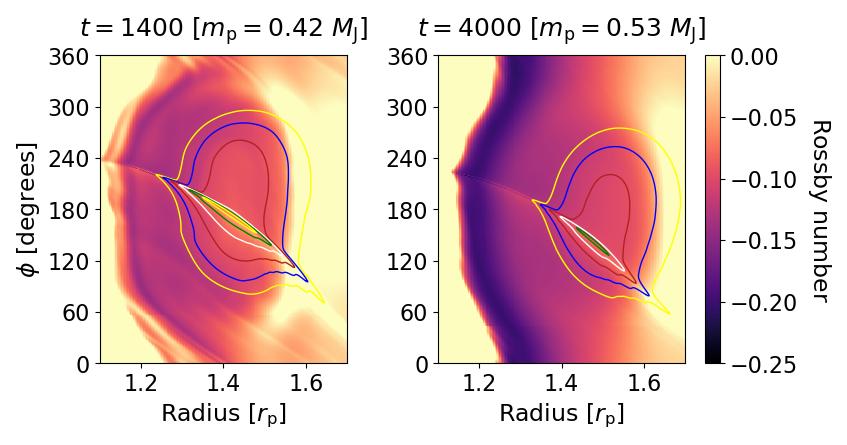}
\caption{Gas density (\textit{top panels}) and Rossby number evolution (\textit{bottom panels}) in the highest-mass case ($A = 0.167$) with $h = 0.08$ and $\nu = 10^{-7}$. Density contours {\cbf(at $\Sigma / \Sigma_0 = 0.8, 0.9, 1.0,$ etc.)} are overlaid on all panels. The initial vortex has a low contrast throughout and is largely unchanged for thousands of orbits due to the weaker shocks from the planet's spiral waves.} 
\label{fig:evolution_h08_a25}
\end{figure}

We tested the dependence on disc mass by running an additional simulation with a $70\%$ lower disc mass and $A = 0.25$ that yields a planet mass of $M_\mathrm{p,~3000} = 0.50~M_\mathrm{Jup}$, a value that lies in-between the second and third highest planet masses from our parameter study. We found that this case yields a low-contrast vortex (see Figure~\ref{fig:evolution_h08_a25}) that closely resembles the case with $h = 0.06$ and a low fixed-mass planet that was not part of our main parameter study, largely because they are both less affected by the planet's spiral waves.

Like the fixed-mass case, the initial vortex that forms is low-contrast and long-lived, surviving until at least $t = 7500~T_\mathrm{p}$ at which point we stop the simulation. The initial vortex forms at $t = 700~T_\mathrm{p}$ and the RWI only grows until $t = 1500~T_\mathrm{p}$ when the vortex reaches a peak azimuthal density contrast of $\approx 2.2$. After this peak, the contrast gradually decays to 1.3 by the end of the simulation (see Figure~\ref{fig:growth_h08}, labelled as $0.3~\Sigma_{0}$), suggesting the vortex can persist much longer. The $m = 1$ dust asymmetry lifetime of over 6800 orbits is already nearly twice as long as the lifetimes in either of the two higher-mass $h = 0.08$ cases from our parameter study, the latter of which has the most similar final planet mass.

Furthermore, we showed this long-lived outcome occurs independent of planet mass in the sub-thermal mass regime by running two additional simulations, a higher-mass case with $A = 1.0$ and $M_\mathrm{p,~3000} = 0.73~M_\mathrm{Jup}$, and a lower-mass case with $A = 0.1$ and $M_\mathrm{p,~3000} = 0.36~M_\mathrm{Jup}$. The vortex in the higher-mass case also survives until at least $t = 7500~T_\mathrm{p}$. Although the vortex in the lower-mass case spreads into a ring slightly earlier at $t = 6600~T_\mathrm{p}$ likely due to the vortex being more elongated, a later-generation vortex forms within ten orbits because of the low separation between the maxima ($r_\mathrm{pressure} - r_\mathrm{crit} < 3.25~H_0$). The first two later-generation vortices extend the total lifetime beyond $t = 7500~T_\mathrm{p}$, demonstrating that later-generation vortices can extend asymmetry lifetimes in low-mass planet cases with $h = 0.08$ as well.





\subsection{Lower Aspect Ratio ($h = 0.04$)} \label{sec:small-h}

We find that the vortices in thinner discs with an aspect ratio of $h = 0.04$ are the shortest-lived among the lower-viscosity cases in our parameter study. Whereas all of the initial $h = 0.06$ gas vortices survive at least 1000 orbits, only the highest-mass $h = 0.04$ case generates an initial vortex that survives nearly that long even though it is more compact than any of the $h = 0.06$ initial vortices. We suspect these lifetimes are the shortest due to the planet's spiral waves having the strongest shocks. Despite these short lifetimes, these cases are prone to forming many successive later-generation vortices because the pressure bump stays relatively close to the planet at the early times in the simulations. 

All four $h = 0.04$ cases in our parameter study have final planet masses above the relatively low thermal mass of $M_\mathrm{th} = 0.19~M_\mathrm{Jup}$. As a result, every planet opens up a relatively deep gap rather quickly. Overall, the lowest-mass case has the longest total dust asymmetry lifetime of 2700~$T_\mathrm{p}$. The lifetimes decline until the second-highest mass case, which has a vortex that survives less than half as long. In the highest-mass case, the lifetime begins to increase again due to the initial gas vortex being stronger from the start.


\subsection{Simulations with High Viscosity ($\nu = 10^{-6}$)} \label{sec:high-visc}

We find that vortices in discs with a much higher viscosity of $\nu = 10^{-6}$ are much shorter-lived. While all of the lower-viscosity cases in the $h = 0.06$ cases have lifetimes above 2000 orbits, none of the higher-viscosity cases have lifetimes more than 1100 orbits. Meanwhile, the longest lifetime among the $h = 0.08$ cases is about 1500 orbits. Moreover, the lowest-mass case ($A = 0.02$: $M_\mathrm{p,~3000} = 0.37~M_\mathrm{Jup}$) never produces a vortex. These lifetimes are a stark contrast from the lower-viscosity $h = 0.08$ cases where a vortex is still present at the end of every simulation ($t > 7500~T_\mathrm{p}$) regardless of planet mass or disc mass.

Like with a lower viscosity, later-generation vortices can still form due to the planet's spiral waves generating a vortensity drop interior to the location of the pressure bump. For both $h = 0.06$ and $h = 0.08$, they follow the same pattern and initially form regular later-generation vortices followed by interior vortices. Unlike the lower-viscosity cases, however, the interior vortices are typically rather short-lived due to the higher viscosity smoothing out the vortensity gradients at the interior and exterior sides of the outer gap edge. We suspect the lowest-mass $h = 0.08$ planet will never trigger a vortex on its growth track because the final gap depth of $\delta_\mathrm{gap} = 4.7$ is much larger than the gap depths of $\delta_\mathrm{gap} < 3$ needed to trigger vortices in the other cases.


\section{Synthetic Images} \label{sec:synthetic}

\begin{figure*} 
\centering
\includegraphics[width=0.88\textwidth]{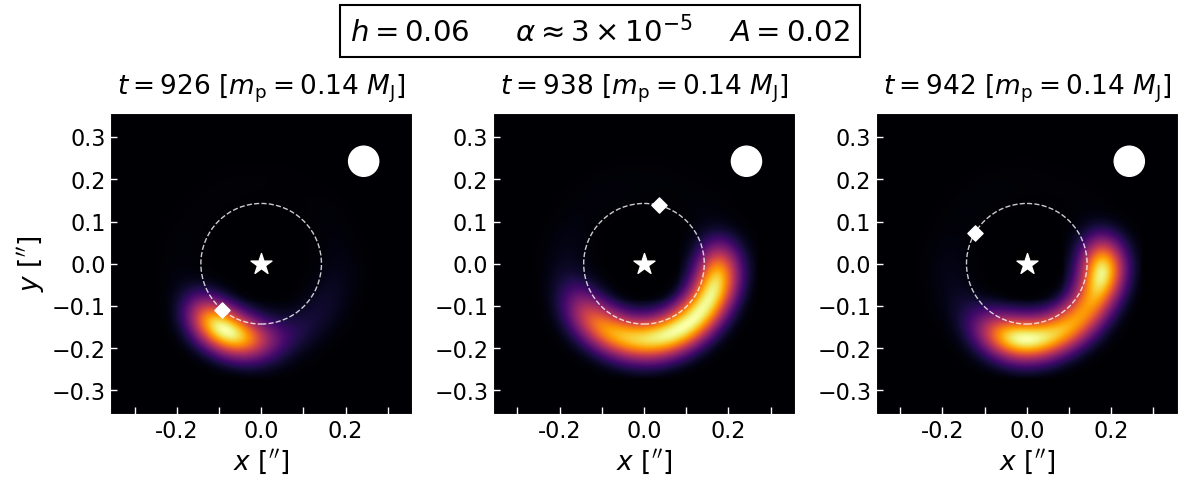}
\includegraphics[width=0.096\textwidth]{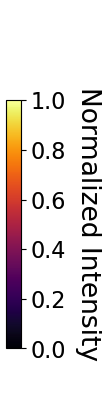}
\caption{Synthetic images with $h = 0.06$ with the dust in the outer disc largely depleted and a beam diameter of $0.284^{\prime \prime}$ (1.0 $r_\mathrm{p}$, or 20 AU). The dust cycles through appearing concentrated (\textit{left}) or elongated (\textit{center}), and can also have two peaks (\textit{right}) over a short span of 20 orbits as it circulates around the vortex. {\cbf The secondary peak in the two-peak case can also be less pronounced than the primary, similar to what has been observed in HD 142527 \citep{boehler21}.}} 
\label{fig:h06-synthetic}
\end{figure*}

\begin{figure*} 
\centering
\includegraphics[width=0.88\textwidth]{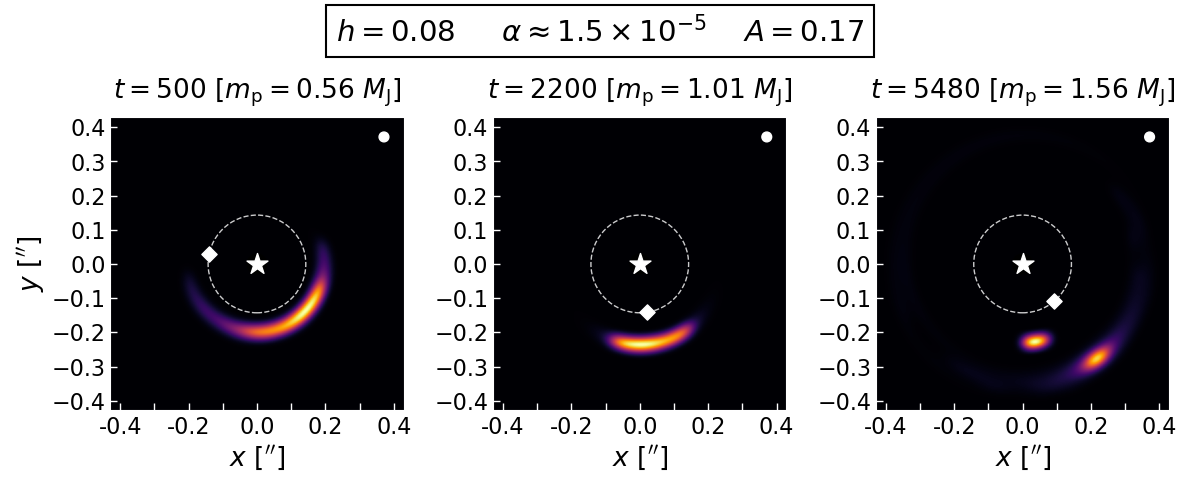}
\includegraphics[width=0.096\textwidth]{figures/intensity_colorbar.png}
\caption{Synthetic images with $h = 0.08$ with a beam diameter of $0.057^{\prime \prime}$ (0.2 $r_\mathrm{p}$, or 4 AU). An elongated vortex (\textit{left}), a compact vortex (\textit{center}), and two vortices after the initial vortex has migrated outwards (\textit{right}). The left panel resembles the crescent in HD 135344B. The right panel resembles MWC 758, while the subsequent stage after the inner vortex disappears resembles Oph IRS 48.} 
\label{fig:h08-synthetic}
\end{figure*}

\subsection{Method} \label{sec:synthetic-method}

We generate synthetic intensity images by computing the flux from individual grid cells that corresponds to the dust surface density in those cells, focusing on the vicinity of the outer gap edge. We then convolve the overall image with an artificial circular beam to illustrate the effects of having finite resolution in observations. The $F$ at a wavelength $\lambda$ in a given cell is
\begin{equation} \label{eqn:flux}
F_\mathrm{\lambda}(r, \phi) = I_\mathrm{\lambda}(r, \phi) \frac{r \delta r \delta \phi}{d^2},
\end{equation}
where  $\delta r$ and $\delta \phi$ are the radial and azimuthal dimensions of the cell and the intensity $I_\mathrm{\lambda}$ is calculated as
\begin{equation} \label{eqn:intensity}
I_\mathrm{\lambda}(r, \phi) = B_{\lambda}(T)[1 - e^{-\tau_{\lambda}}].
\end{equation}
In that expression, $B_{\lambda}(T)$ is the spectral radiance following Planck's law, and  the optical depth $\tau$ is 
\begin{equation} \label{eqn:tau}
\tau_{\lambda} = \Sigma_\mathrm{dust}(r, \phi; s) \kappa_{\nu},
\end{equation}
where $\kappa_{\lambda}$ is the opacity of a dust grain of a given size when observed at a given wavelength. These values for the opacity are derived from the Jena database\footnote{http://www.astro.uni-jena.de/Laboratory/Database/databases.html}, assuming Mie theory applies \citep[$2 \pi s \approx \lambda$, see][]{bohren83} and dust grains that are magnesium-iron silicates \citep{jaeger94, dorschner95}.

We calculate the images at $\lambda = 0.87$ mm, which lies within ALMA Band 7. In order to add a physical scale the hydrodynamics simulations, we set the separation between the star and the planet to $r_\mathrm{p} = 20$ AU. We incorporate a temperature profile of $T = T_0 (r / r_\mathrm{p})^{-1}$ that is consistent with the scale height and flaring index from our hydrodynamic simulations assuming a mean molecular weight of $\mu = 2.34$. We fix the star's temperature to that of the Sun, $T_\mathrm{\odot} = 5770$ K, and view the disc face-on with an inclination of $i = 0^{\circ}$.

We convolve the flux image with different beam diameters, either $0.057^{\prime \prime}$ or $0.284^{\prime \prime}$, which correspond to scales of 4 AU and 20 AU in the disc.
In order to focus on the vortex at the outer edge of the gap, we filter out any contributions from the dust co-orbital with the planet and further interior ($r < 1.08~r_\mathrm{p}$) to avoid contaminating the outer disc with the larger artificial beam diameters. 

\subsection{Results} \label{sec:synthetic-results}

Unlike in \cite{hammer19} where we found that elongated gas vortices always appear elongated in the dust, we find that the elongated gas vortices in this study can appear as either compact or elongated as the dust circulates around the vortex. Like in \cite{hammer19}, we also find that the dust typically collects in off-center peaks as a result of that circulation. 

Figure~\ref{fig:h06-synthetic} shows three distinct example synthetic images of vortices with $h = 0.06$ while the gas vortex is alive. To calculate these images, we initialized the dust at $t = 500~T_\mathrm{p}$ and incorporated an exponential cutoff of $e^{-(r/r_\mathrm{cut})^5}$ to the initial dust surface density profile. With $r_\mathrm{cut} = 1.6~r_\mathrm{p}$, the dust in the outer disc depletes by $>80\%$ by $900~T_\mathrm{p}$ and fully depletes by $1000~T_\mathrm{p}$. This dust distribution better resembles discs such as HD 135344B and RY Lupus where no dust is observed at any distance beyond the asymmetry, either because the dust supply is depleted or because the dust growth is limited at those distances. 

With a limited dust supply, we find that the dust largely collects in a radially narrow clump. Near the azimuthal edges of the vortex, this clump is also azimuthally compact, creating a compact signature. In the middle of the vortex, however, it commonly spreads out in azimuth, creating an elongated signature that on occasion can have two peaks.\footnote{Although we had also found two-peak signatures in the $m = 1$ phase in \cite{hammer19}, those signatures were influenced by our use of the terminal velocity approximation, which breaks down near spiral shocks \citep{lovascio19}. Nonetheless, we find that same signature in this study without the use of that approximation.} We ran a comparison without the exponential cutoff and found that the two-peak signature only appears with the cutoff. As with the cutoff, the vortex can also appear compact or elongated without the cutoff. It may be more difficult to measure the azimuthal extent without the cutoff, however, due to the dust piling up just exterior to the vortex and creating a ring-like feature that blends with the vortex with too low of a resolution.

In general, we find that elongated vortices in discs with different aspect ratios and associated with different-mass planets have similar appearances due to their similar azimuthal extents. In the cases with $h = 0.08$, however, the appearance of the vortex can vary significantly after it becomes more compact. Figure~\ref{fig:h08-synthetic} compares three different stages of the highest-mass $h = 0.08$ case, highlighting when the vortex is elongated, when the vortex is compact, and when two vortices are present. Even with the original vortex located at the outer pressure bump, we find that the inner later-generation vortex can still collect dust.

\subsection{Origin of vortex signatures} \label{sec:origin}

Despite the same two signatures -- an elongated azimuthal extent and an off-center peak  -- being present as in our previous work, the explanation for these signatures is slightly different. 

In \cite{hammer19}, we had simulated a Jupiter-mass planet grown over 750 orbits on a prescribed sine-squared profile and found that the resulting vortex lacked a series of smooth elliptical surface density contours across a wide region in its center. Instead, the vortex had a highly disrupted pattern of contours in large part due to the planet's spiral density waves repeatedly passing through and shocking the vortex. These waves also kept the dust preferentially towards the front side of the vortex even as it circulated around.

On the contrary with the more realistic planet growth in this work, we find that the vortices in our study largely maintain their series of elliptical contours in their centers. In the simulations with higher-mass planets, we attribute this behavior to the planet generating a stronger vortex from the start. The vortex is stronger because the planet is initially growing at a faster rate compared to the planet in our previous study. In the simulations with lower-mass planets, we attribute these differences to the planet's weaker spiral waves not affecting the vortex as much. 

Nonetheless, even without the planet's spiral waves having a strong influence on the dust dynamics, the dust peak still circulates around the vortex. Unlike with the stronger waves, the dust circulates around the entire vortex, instead of staying towards the front. This circulation can create compact or elongated signatures.



\section{Caveats} \label{sec:caveats}

\begin{table*}
\caption{Likelihoods of observing vortices adjacent to gaps from \protect\cite{long18}, none of which have asymmetries. Only those with planet masses $M_\mathrm{p} / M_\mathrm{\bigstar} \ge 0.3~M_\mathrm{Jup} / M_\mathrm{\odot}$ as fit by \protect\cite{lodato19} are included. Stellar parameters from \protect\cite{long19}.}
\label{table:fractions}
\begin{threeparttable}
\begin{tabular}{ c || c c c | c c c || c c c | c c c}
  Name & $M_\mathrm{p}$ & $M_\mathrm{\bigstar}$ & $M_\mathrm{p} / M_\mathrm{\bigstar}$ &$R_\mathrm{p}$ & $T_\mathrm{p}$ & $t_\mathrm{\bigstar}$ & $f_\mathrm{1000}$ & $f_\mathrm{2000}$ & $f_\mathrm{4000}$ & $f_\mathrm{c,1000}$ & $f_\mathrm{c,2000}$ & $f_\mathrm{c,4000}$\\
            & ($M_\mathrm{Jup}$) & ($M_\mathrm{\odot}$) & ($M_\mathrm{Jup} / M_\mathrm{\odot}$) & (AU) & (yr) & (Myr) \\
   \hline
        \hline
  DS Tau & 5.6 & 0.58 & 9.60 & 32.9 & 248 & $4.8^{+4.8}_{-3.2}$ & 0.052 & 0.103 & 0.207 & 0.124 & 0.248 & 0.496\\
  FT Tau & 0.15 & 0.34 & 0.44 & 24.8 & 212 & $3.2^{+3.2}_{-1.6}$ & 0.066 & 0.132 & 0.264 & 0.106 & 0.212 & 0.424\\
  MWC 480 & 1.3 & 1.91 & 0.68 & 73.4 & 455 & $6.9^{+5.1}_{-5.8}$ & 0.066 & 0.132 & 0.264 & 0.228 & 0.455 & 0.911 \\
  DL Tau & 0.33 & 0.98 & 0.34 & 88.9 & 847 & $3.5^{+2.8}_{-1.6}$ & 0.242 & 0.484 & 0.968 & 0.423 & 0.847 & 1.000 \\
  CI Tau & 0.36 & 0.89 & 0.40 & 48.4 & 357 & $2.5^{+2.0}_{-1.1}$ & 0.143 & 0.285 & 0.570 & 0.178 & 0.356 & 0.713 \\
  CI Tau & 0.37 & 0.89 & 0.42 & 119.0 & 1376 & $2.5^{+2.0}_{-1.1}$ & 0.550 & 1.000 & 1.000 & 0.688 & 1.000 & 1.000 \\
     \hline
         Total (out of 6) & & & & & & & 1.11 & 2.14 & 3.27 & 1.75 & 3.12 & 4.54
\end{tabular}
\begin{tablenotes}
   \item \textit{Columns:} (1) Disc name, (2) Planet mass $M_\mathrm{p}$, (3) Stellar mass $M_\mathrm{\bigstar}$, (4) Planet-to-star mass ratio $M_\mathrm{p} / M_\mathrm{\bigstar}$, (5) Planet separation $R_\mathrm{p}$, (6) Planet orbital period $T_\mathrm{p}$, (7) Stellar age, (8) (9) (10) Fractions of stellar age where vortex would be present assuming lifetimes of 1000, 2000, and 4000~$T_\mathrm{p}$, (11) (12) (13) Fractions of cluster age where vortex would be present assuming a Taurus cluster age of 2 Myr.
   \item  \textit{Bottom row:} Total expected number of vortices with each assumed vortex lifetime and stellar age or cluster age. There would be at least two vortices expected with lifetimes $\ge 2000~T_\mathrm{p}$ that are characteristic in discs with very low viscosities ($\alpha \sim 3 \times 10^{-5}$), which is inconsistent with observations that all of these discs are symmetric.
    \end{tablenotes}
\end{threeparttable}
\end{table*}

\subsection{Effects of dust feedback} \label{sec:dust-feedback}

As vortices can quickly accumulate high amounts of dust due to the fast drift speeds of large grains, the feedback effects of the dust on the evolution of the gas quickly become relevant. 
Although the earliest 2-D studies of vortices that incorporated these effects showed that feedback can reduce vortex lifetimes \citep{fu14b}, more recent 3-D work has demonstrated that feedback has much less of an effect on the overall structure of vortices away from the midplane and thereby should not affect vortex lifetimes \citep{lyra18}.

As such, the primary effect feedback should have on vortices is to alter their appearance in the dust. Previous studies have shown that with sufficiently high dust-to-gas ratios, feedback concentrates dust into more compact clumps or produces multiple clumps in the same vortex \citep{fu14b, crnkovic15, miranda17}. In our preliminary studies of elongated vortices with dust feedback, we found the same type of clumped dust signatures, which do not match the appearance of elongated crescent-shaped dust asymmetries in observations such as HD 135344B \citep{vanDerMarel16b, cazzoletti18}. It is possible that feedback might not affect observed vortices because they do not have dust-to-gas ratios of order unity, either due to the vortex having a low supply of large dust grains from the outer disc where grain growth times are slow, or from the formation of planetesimals depleting the vortex's supply of dust. We will present the effects of dust feedback on elongated vortices in more detail in a forthcoming publication.



\subsection{Effects of self-gravity} \label{sec:self-gravity}

Unlike simpler hydrodynamic simulations of discs that are scale free, the results of our study are dependent on the disc mass because of the role it plays in governing the growth of the planet. Moreover, self-gravity should affect our results at higher disc masses, in particular with the disc masses in our parameter study. On the other hand, the simulations we conducted with lower disc masses should not be affected by self-gravity.

{\cbf We tested whether self-gravity can prevent the unimpeded RWI growth and outward vortex migration scenario from the $h = 0.08$ cases in our parameter study by replicating the highest-mass case ($A = 0.67$) while adding self-gravity using the FARGO-ADSG code \citep{FARGO-ADSG}. Although these $h = 0.08$ cases initially have a relatively high $Q = 22.5$ (with $\Sigma = 1.0~\Sigma_0$ and $r = 1.4~r_\mathrm{p}$), where $Q$ is the Toomre Q parameter \citep{toomre64}, the increase in density in the vortex causes a drop to $Q = 7.6$ (with $\Sigma = 2.0~\Sigma_0$ and $r = 1.7~r_\mathrm{p}$).  \cite{lovelace13} have shown that self-gravity can prevent the disc from becoming unstable to the RWI in idealized circumstances when $Q < (\pi/ 2)(H/r)^{-1}$, where this critical value is $Q = 19.6$ in our $h = 0.08$ cases.  As such, we find that self-gravity eventually stops the growth of the RWI, preventing it from shrinking or eventually migrating outwards, consistent with expectations from previous work \citep{mkl11, mkl12b, zhu16}. Instead, the vortex behavior matches that of the $h = 0.08$ cases with a lower disc mass, surviving for several thousands of orbits due to the weaker shock from the planet's spiral waves. Therefore, vortices in low-viscosity discs with $h = 0.08$ can still be long-lived at high disc masses even with the effects of self-gravity.}


{\cbf Because self-gravity can prevent the unimpeded growth scenario from occurring in sufficiently massive discs,} we tested whether lower-mass discs can also produce that growth and outward vortex migration with an additional simulation that has a $40\%$ lower disc mass and a larger aspect ratio of $h = 0.1$. The planet accretes with $A = 0.1$ and grows to $M_\mathrm{p,~3000} = 0.85~M_\mathrm{Jup}$. We found that this case reproduces the three main features from our parameter study, namely unimpeded RWI growth, outward migration of the vortex, and the formation of a later-generation vortex. Unlike the $h = 0.08$ cases from our parameter study, this case also maintains a value of $Q \times (H/r) > \pi / 2$ throughout the simulation, indicating that this scenario can still occur in discs not massive enough for self-gravity to affect the outcome. 




\subsection{Effects of planet migration} \label{sec:planet-migration}

{\cbf One simplified feature of our simulations is that the planets remain on fixed orbits even though an actual planet would naturally migrate due to the uneven torques exerted on the planet by the disc material in its vicinity \citep[see review by][]{nelson18}. This migration of the planet should affect how quickly it opens up a gap and the shape of the gap, factors that affect both how much the planet can grow and the properties of any resulting vortices that form at the gap edges.

We tested whether the planet migrating could inhibit the formation of later-generation vortices by running two additional simulations with $h = 0.06$ and $\nu = 10^{-7}$, one with $A = 0.04$ and our fiducial disc mass that yields a planet with $M_\mathrm{p,~3000} = 0.57~M_\mathrm{Jup}$, and another with $A = 0.1$ and a $70\%$ lower disc mass that yields a planet with $M_\mathrm{p,~3000} = 0.34~M_\mathrm{Jup}$. In both simulations, the planet initially migrates inward (Type I) before speeding up as a partial gap is opened (Type III). After the gap empties by a sufficient amount, however, the planet stops migrating inwards and begins to migrate outwards at a very slow, near-negligible rate due to the asymmetry in the gap structure, matching the pattern found by \cite{lega20}. Because the planet largely stops migrating in the third stage, it can still generate later-generation vortices, albeit potentially with a different type of signature. We will present the consequences of planet migration on elongated vortices in depth in a future study.

}


\section{Applications to Observations} \label{sec:applications}

\subsection{Why don't more discs contain dust asymmetries?} \label{sec:low}

With the long vortex lifetimes we have found in this study, why aren't more planetary gap candidates associated with dust asymmetries? Assuming each gap was opened by a planet, we can estimate the likelihood of observing an asymmetry at specific gaps by assuming a vortex of a given lifetime was induced by a planet in each gap and then comparing the vortex lifetime to the age of the system. We apply this method to the sample of gaps in Taurus from \citep{long18}, where the cluster age is 1-2 Myr and discs with substructures have a slightly higher median age of 3.2 Myr \citep{long19}. We only include planet candidates where the planet-to-star mass ratio exceeds Saturn-to-solar mass, according to a simple model by \cite{lodato19}.\footnote{Although we have shown that planets with even lower masses can trigger long-lived vortices, the fitted planet masses from \cite{lodato19} would likely be overestimates by a factor of $\sim3$ (e.g. \citealp{DSHARP-VII}) if these discs indeed have very low viscosities.} None of these six gaps are associated with an asymmetry.

In Table~\ref{table:fractions}, we compare vortex lifetimes of 1000, 2000, and $4000~T_\mathrm{p}$ to both the cluster age of 2 Myr and the stellar ages, the former of which is more reliable due to the high uncertainties associated with each individual stellar age. We find that with assumed lifetimes of $1000~T_\mathrm{p}$ --- which is characteristic in a disc with a higher viscosity near $\alpha \sim 2 \times 10^{-4}$ --- we would expect to find one to two vortices between all six systems based on the fraction of the system lifetime when a vortex would be present. That number should be an underestimate given both that planets are not likely to form when the system is very young and also that self-gravity could prevent vortices from arising if a planet formed in a disc young enough to still be very massive. 

We would expect to see at least two vortices with longer lifetimes of $2000~T_\mathrm{p}$ and at least three with lifetimes of $4000~T_\mathrm{p}$. With lifetimes of $2000~T_\mathrm{p}$ and the cluster age in particular, all six gaps have at least a one-fifth probability ($f_\mathrm{c, 2000} > 0.2$) of being accompanied by a vortex, which suggests there should be a higher amount of asymmetries than the zero that are observed. As these longer lifetimes are more characteristic of vortices in discs with lower viscosities of $\alpha \sim 2 \times 10^{-5}$, we assert that this sample of discs in Taurus should have higher viscosities in order to explain the lack of asymmetries in these systems. It could also be the case that the associated planets formed too early in the disc lifetime for vortices to arise, or that these gaps are not associated with planets.

\subsection{Preference for vortices at large separations} \label{sec:large-separations}

One peculiarity of known crescent-shaped asymmetries is that they are preferentially located in the outer disc, even though existing observations \citep[e.g. the DSHARP survey:][]{DSHARP-I} are capable of resolving such features at 20 AU or less. Nearly all of these features are located at $\ge 50$ AU -- e.g. in IRS~48 \citep{vanDerMarel13}, MWC~758 \citep{marino15}, HD~135344B \citep{vanDerMarel16b}, RY Lupus \citep{ansdell16}, and T4 \citep{pascucci16}. The inner asymmetry in HD 163296 at 4 AU is the only known exception thus far \citep{DSHARP-IX}.

It is natural to expect vortices to be longer-lived in the outer disc where the Keplerian orbital timescales  are much longer. Due to flaring though, the outer region of a disc should also have a larger aspect ratio \citep{chiang97}. \cite{fu14a} had previously found, however, that an aspect ratio of $H/r = 0.06$ maximized vortex lifetimes with $5~M_\mathrm{Jup}$ planets and that larger aspect ratios lead to much shorter lifetimes. They argue that dependence was due to the initial bump at the outer gap edge being much wider relative to the amplitude due to the larger scale height, thereby creating a weaker vortex \citep{ono16}. Although we also find that the vortex initially has a much lower peak density compared to lower aspect ratio cases, we do not find that this lower density shortens the lifetime of the vortex.

Our findings that (1) low-mass planets induce long-lived vortices with $H/r = 0.06$, and that (2) every case with $H/r \ge 0.08$ has a vortex alive at the end of the simulations lends support to ALMA preferentially finding nearly all vortex candidates in the outer disc at $\ge 50$ AU, where these discs should have large aspect ratios.



\subsection{Location of dust trap in Oph IRS 48} \label{sec:irs48}

The disc around Oph IRS 48 contains an asymmetric dust trap at 61 AU \citep{vanDerMarel15b} even though CO observations found that the gas cavity in the inner part of the disc only extends out to about 20 AU, three times closer in than the dust asymmetry itself \citep{vanDerMarel13, bruderer14}. Previous work has attempted to model this system with a planet triggering a vortex that in turn triggers another vortex at a larger separation that survives longer than the first \citep{lobogomes15}. More recent work has modeled this system with an undetected companion star driving an eccentricity and an associated instability at the inner edge of the cavity \citep{ragusa17, calcino19}. Alternatively, we suggest that the initial vortex migrating outwards -- as it does in the large aspect ratio ($h \ge 0.08$) cases with high disc masses -- could potentially explain that wide separation. 

In those cases, the initial vortex migrates outwards as a result of residing in the middle of a positive pressure gradient after the vortex itself causes the initial pressure bump to deplete by becoming more compact. After a second-generation vortex forms and dissipates, the original vortex is left as the only major dust feature in the outer disc at a location of $r > 2.0~r_\mathrm{p}$. Even though the vortex ends up far from the planet, the outer edge of the gap (or cavity) in the gas is still much closer to the planet (see Figure~\ref{fig:avg_density_h08}). The separation in Oph IRS 48 may be even larger than in our simulations because it has a larger aspect ratio. \cite{bruderer14} fit the aspect ratio to $h = 0.14$ at the location of the asymmetry. 
This large aspect ratio falls safely in the $h \ge 0.08$ range in which the vortex can migrate outwards in our parameter study, and also makes it less likely for disc self-gravity to affect this scenario.

\subsection{Multiple asymmetries in MWC 758} \label{sec:mwc758}

The disc around MWC 758 contains two asymmetries at about 55 AU and 80 AU \citep{boehler18}. Previous work has attempted to model these features with two gas giant planets \citep{baruteau19}, one interior to the inner asymmetry and the other exterior to the outer asymmetry that is also responsible for the spirals in the system \citep{ren20}. Recently, it has also been suggested that the inner planet could produce the spirals if it has an eccentric orbit \citep{calcino20}. Like with Oph IRS 48, the scenario of an initial vortex migrating outwards from our parameter study could also potentially explain the multiple asymmetries.

Whereas Oph IRS 48 resembles the time after the later-generation vortex has dissipated, the MWC 758 system could instead resemble the third panel of Figure~\ref{fig:h08-synthetic} with both the initial vortex (the outer asymmetry) and the later-generation vortex (the inner asymmetry) present. As this disc has been modelled to have $h = 0.088$ \citep{boehler18, baruteau19}, the properties of this system may be consistent with the parameters needed for this scenario to occur. Given that \cite{ren20} modeled the outer planet to be rather far from the outer asymmetry at 172 AU, we suggest that an inner planet could instead potentially be responsible for both asymmetries. Meanwhile, an outer planet could still be responsible for the spiral features.


\section{Conclusions} \label{sec:conclusions}

\subsection{Summary} \label{sec:summary}

In this study, we investigated the lifetimes of dust asymmetries associated with vortices over a  range of planet masses, disc aspect ratios, and disc viscosities. Our key innovation in the methodology of the study was to have the planet accrete its gas from the disc instead of prescribing this growth. Just like with prescribed growth, we find that our accreting planets generally trigger vortices that are elongated. {\cbf Although these vortices are generally weaker compared to compact vortices in terms of their density contrasts, we find that they can still be long-lived and exhibit a variety of different dust signatures.}

{\cbf \textit{Which planets trigger longer-lived vortices: low-mass or high-mass?}} Like planets at Jupiter-mass and above, we find that lower-mass planets can also generate long-lived dust asymmetries at the outer gap edge that frequently have even longer lifetimes than the asymmetries associated with higher-mass planets in discs with the same parameters. Regardless of planet mass, these asymmetries require very low viscosity of $\nu = 10^{-7}$ (equivalent to $\alpha < 10^{-4}$) to survive more than 1500 planet orbits. With an aspect ratio of $H/r = 0.06$, we find that dust asymmetry lifetimes exceed 4500 orbits for low-mass planets ($M_\mathrm{p} < 0.4~M_\mathrm{Jup}$) below the classical gap-opening mass. Meanwhile with $H/r = 0.08$, there are still asymmetries present at the end of all of our simulations --- all of which are run out to at least 7500 orbits (corresponding to vortex lifetimes above 6000 orbits) --- regardless of planet mass or disc mass.

{\cbf \textit{Why are the vortices elongated, and do they appear elongated in the dust?}} We characterize the vortices in our study as elongated because of their wide azimuthal extents and their less negative minimum Rossby numbers of $\mathrm{Ro} > -0.15$, characteristics that are captured by the GNG model from \citealp{GNG}. These $m = 1$ vortices end up elongated because the initial $m \ge 1$ set of vortices that form also have $\mathrm{Ro} > -0.15$ and also lack a clear vorticity minimum at the center of the vortex. We find that vortices that are elongated in the gas can appear either elongated or compact as the dust circulates around the vortex, and commonly have off-center peaks due to the dust not settling at the center of the vortex.

{\cbf \textit{Why do planet-induced vortices dissipate?}} We demonstrate that a combination of the planet's spiral waves and the disc's viscosity are responsible for elongated vortices spreading into rings. The shocks from the planet's spiral waves can disrupt the vortex's elliptical density structure and vortical motion. The viscosity halts the growth of the Rossby Wave instability after about $10^3$ orbits in most cases, preventing the vortex from continuing to develop a stronger density maximum or vorticity minimum. Once this growth ends, the initial vortex becomes susceptible to being disrupted by shocks and does not survive more than a few hundred orbits in the cases where the shocks are strong. As such, the initial gas vortices with $H/r = 0.04$ are the shortest-lived because they experience the strongest shocks. On the contrary, the initial vortices in the $H/r = 0.08$ cases can survive for thousands of orbits after the RWI stops growing --- more than twice as long as the initial $H/r = 0.06$ vortices from our parameter study --- because they experience the weakest shocks.

{\cbf \textit{How and when can vortices re-form?}} Despite the spiral waves and viscosity destroying the initial gas vortices in the $H/r = 0.06$ cases rather quickly, the cases with lower-mass planets still sustain long-lived asymmetries in the dust because they can spawn later-generation vortices to replace the original. Whether another vortex can form depends on the locations of the peaks in two related radial profiles: the pressure bump and the RWI critical function. By the time the initial gas vortex spreads into a ring, the locations of these two maxima have both migrated outwards compared to where they were when the initial vortex formed. More importantly, the separation between the two maxima has increased as well. We find that these maxima need to stay relatively close ($r_\mathrm{pressure} - r_\mathrm{crit} < 3.25~H_0$) to form a later-generation vortex. Given that the location of the pressure bump moves farther from the planet with an increasing planet mass while the location of the RWI critical function maximum appears to be largely independent of planet mass, later-generation vortices preferentially arise with lower-mass planets.

{\cbf \textit{What other signatures can planet-induced vortices have?}} The only cases where a non-zero viscosity does not halt the growth of the RWI are the ones with large disc aspect ratios ($H/r \ge 0.08$) and the high disc mass in our parameter study. The RWI continues to grow in these cases because the gap remains shallow and the planet keeps growing at a substantial rate. Eventually, the vortex grows enough to transition from elongated (where $\mathrm{Ro} > -0.15$) to compact (where $\mathrm{Ro} < -0.15$ and the vortex is well-described by the Gaussian model from \citealp{surville15}). At this point, the vortex sheds angular momentum through its own spiral waves and ultimately migrates away from the planet to $r > 2.0~r_\mathrm{p}$. During this phase, a relatively long-lived later-generation vortex can arise at the location of the original before ultimately dissipating and leaving the initial vortex alone in the outer disc at a much farther separation from the planet.


\subsection{Implications} \label{sec:implications}

With dust asymmetries being longer-lived in regions of discs with larger aspect ratios, it is even more natural to expect to find asymmetries at large separations from their stars, where orbital timescales are longer and aspect ratios are larger. This result differs from the optimal intermediate aspect ratio that \cite{fu14a} found in their study of planets with a higher fixed mass.

With the long dust asymmetry lifetimes at low disc viscosities ($\alpha \sim 2 \times 10^{-5}$) and the large orbital separations of the gap population from ALMA observations, it would be natural to expect the fraction of discs containing asymmetries to be larger than the 20 to $25\%$ that is observed. Planets at semimajor axes of $\sim30$ to $60$ AU would only need to induce vortices that survive for about 2000 planet orbits to last 0.5 to 1 Myr, already a significant fraction of the disc lifetime in young to medium-aged discs. We have demonstrated those lifetimes are typical when $H/r \ge 0.06$, and that they can underestimate the lifetime by a factor of two or more with planets of Saturn-mass and below or in regions of discs with larger aspect ratios. In particular, we showed in the Taurus sample of discs from \cite{long18} that there are enough high-enough mass planet candidates at distant-enough separations from their stars to expect to find multiple asymmetries, even though none were observed. We propose this discrepancy is a consequence of these discs having higher viscosities ($\alpha \ge 10^{-4}$).

These discs having high values of $\alpha$ would support MHD simulations that found discs to have effective values of $\alpha$ higher than the lower set of values of $\alpha$ in our parameter study, even with non-ideal effects and no prescribed viscosity \citep[e.g.][]{zhu14}. Another possible outcome with high viscosities is that lower-mass planets might not trigger vortices at all, which is also a more natural outcome with slow prescribed planet growth times \citep{hallam20}. As such, we might observe so few asymmetries because planets in discs with these higher viscosities never triggered vortices.

The scenario we found where the vortex migrates outwards is particularly interesting because it creates two qualitatively different outcomes from a usual planet-induced vortex in the outer disc. First, it makes it possible for there to be two asymmetries in the outer region of a disc with only one planet. Second, the outward migration could then leave the initial vortex at a much larger separation from the planet compared to normal. The first outcome could potentially explain the two clumps observed in MWC 758, while the second outcome could potentially explain the large separation between the planet candidate and the asymmetric dust trap in Oph IRS 48. This scenario can be inhibited by disc masses that are too low or too high. Disc masses that are too low prevent the vortex from migrating outwards as a result of the planet gap depth being too large. Disc masses that are too high {\cbf are} susceptible to self-gravity preventing the vortex from growing enough to migrate. As such, this scenario may require both a high aspect ratio ($H/r \ge 0.08$) and an intermediate disc mass.


\section*{Acknowledgements}

The authors would like to thank the referee for helpful comments that improved the manuscript. MH would also like to thank Phil Armitage, Camille Bergez-Casalou, James Cho, Robin Dong, Josh Eisner, Willy Kley, Hui Li, Shengtai Li, Rixin Li, Ya-Ping Li, Feng Long, and Sijme-Jan Paardekooper for helpful one-on-one discussions. Additionally, MH would like to thank Bertram Bitsch for organising the \textit{From protoplanetary discs to planetary systems} workshop where the early stages of this work were first presented, Carlo Manara and Carsten Dominik for hosting me to present this work, and also Phil Armitage, Sijme-Jan Paardekooper, and Kees Dullemond for giving me the opportunity to present this publication virtually. Lastly, MH would like to thank the organisers of \textit{Five years after HL Tau: a new era in planet formation} for setting up an excellent virtual conference where this work was also presented. MH is supported by the NSF Graduate Research Fellowship under Grant No. DGE 1143953 and the NASA Space Grant Graduate Fellowship. MKL is supported by the Ministry of Science and Technology of Taiwan under grant 107-2112-M-001-043-MY3 and the Academia Sinica Career Development Award AS-CDA-110-M06. MH and KMK are supported by NASA Astrophysics Theory Grant NNX17AK59G and by the National Science Foundation under Grant No. AST-1616929. PP acknowledges support provided by the Alexander von Humboldt Foundation in the framework of the Sofja Kovalevskaja Award endowed by the Federal Ministry of Education and Research. The El Gato supercomputer, which is supported by the National Science Foundation under Grant No. 1228509, was used to run all of the simulations in this study. \\

\section*{Data Availability}

The data underlying this article will be shared on reasonable request to the corresponding author.



\bibliography{vortex_bibliography}


\appendix
\section{Resolution tests} \label{sec:resolution}

\subsection{Intermediate Aspect Ratio ($h = 0.06$)} \label{sec:h06_resolution}

We performed a resolution test with the lowest-mass case ($A = 0.02$) from our parameter study with $h = 0.06$ and $\nu = 10^{-7}$ by re-running that case up to $t = 3500~T_\mathrm{p}$ at a higher resolution of $N_\mathrm{r} \times N_\mathrm{\phi} = 2048 \times 3072$. Like with the fiducial resolution, the planet triggers the initial vortex at $t \approx 750~T_\mathrm{p}$ followed by a series of three later-generation vortices. These vortices also coincide with maxima arising in the critical function. Like the corresponding fiducial case, the first two of these vortices are short-lived re-triggers while the last one is a longer-lived interior re-trigger at $3230~T_\mathrm{p}$. This simulation supports that sub-thermal mass planets generate long-lived dust asymmetries and that the vorticity drop in the gap is the mechanism for triggering later-generation vortices.

\subsection{Larger Aspect Ratio ($h = 0.08$)} \label{sec:h08_resolution}

We also performed a resolution test with the highest-mass case ($A  = 0.167$) from our parameter study with $h = 0.08$ and $\nu = 10^{-7}$ by re-running that case up to $t = 3500~T_\mathrm{p}$ at the same higher resolution as above of $N_\mathrm{r} \times N_\mathrm{\phi} = 2048 \times 3072$. Just as with the fiducial resolution, viscosity never stops the growth of the RWI, allowing the vortex to eventually transition from elongated to compact and begin to rapidly shrink. This simulation supports the existence of this pathway for creating different types of vortex structures in a disc.

Lastly, we tested that the $h = 0.08$ vortices are still long-lived at lower disc mass by re-running the intermediate planet-mass case ($A = 0.25$) with a lower initial surface density of $0.3~\Sigma_0$ at a lower resolution of $N_\mathrm{r} \times N_\mathrm{\phi} = 768 \times 1024$. As with the fiducial resolution, the initial vortex survives past 8000 orbits, at which point we stop the simulation. This simulation supports that shocks from the planet's spiral waves do not have a significant effect on the vortex with $h = 0.08$.

\section{Planet growth} \label{sec:h06_planet}

The growth tracks of the planet masses themselves in the $h = 0.06$ low-viscosity fiducial set of cases all follow a similar ``inverse parabola" pattern. Initially, the accretion rate onto the planet increases as the planet becomes more massive and its Hill sphere becomes larger, as demonstrated in the second bottom panel of Figure~\ref{fig:mass}. Once the outer gap edge becomes unstable to the RWI, the accretion rate begins to decrease in spite of the planet continuing to grow in mass and the gap having yet to become significantly depleted. The accretion rate then continues to steadily decrease as the vortex becomes stronger. We interpret this behavior as a sign that some of the gas ends up trapped in the vortex instead of flowing towards the star at the usual viscous rate.

Eventually, the gap becomes depleted enough to also limit the accretion onto the planet. At this point, the mass of the planet begins to asymptote towards a final value. This slowing of the accretion rate occurs at:
\begin{itemize}[leftmargin=1em]
\item ~[$A = 0.50$]:~~~$t \approx ~~600~T_\mathrm{p}$~~~($m_\mathrm{p} \approx 0.75~M_\mathrm{J}$),~~~$\delta_\mathrm{gap} \approx 22.5$
\item ~[$A = 0.17$]:~~~$t \approx ~~750~T_\mathrm{p}$~~~($m_\mathrm{p} \approx 0.49~M_\mathrm{J}$),~~~$\delta_\mathrm{gap} \approx ~~9.2$
\item ~[$A = 0.05$]:~~~$t \approx 1250~T_\mathrm{p}$~~~($m_\mathrm{p} \approx 0.28~M_\mathrm{J}$),~~~$\delta_\mathrm{gap} \approx ~~6.1$
\item ~[$A = 0.02$]:~~~$t \approx 2000~T_\mathrm{p}$~~~($m_\mathrm{p} \approx 0.18~M_\mathrm{J}$),~~~$\delta_\mathrm{gap} \approx ~~5.0$,
\end{itemize}
where the planet masses at which the depletion of the gap limits accretion are different due to the planet clearing out the gap slowly instead of instantaneously.

In general, we note that the accretion rates only scale with the accretion coefficient $A$ at $t = 0$. Afterwards, the accretion rates are much more sensitive to the planet's mass and later on the gap's depletion rate and the presence of vortices themselves. Coincidentally, the competition between the planet's growth amplifying the accretion rates and the gap depth limiting the accretion rates leaves all of the planets in the $h = 0.06$ with similar accretion rates as a function of time once the outer gap edges become unstable to the RWI.

 \end{document}